\pdfoutput=1
\documentclass[10pt,journal,compsoc]{IEEEtran}
\usepackage{booktabs} 
\usepackage{graphicx}
\usepackage{subfigure}
\usepackage{amsthm}
\usepackage{amssymb}
\usepackage{amsmath}
\usepackage{multirow}
\usepackage{enumerate}
\usepackage{amsfonts}
\usepackage{color}
\usepackage{url}
\usepackage{bm}
\usepackage{tabularx}
\usepackage{hyperref}
\usepackage{ragged2e}
\usepackage[justification=centering]{caption}
\usepackage[linesnumbered,commentsnumbered,ruled,vlined]{algorithm2e}
\ifCLASSOPTIONcompsoc
  \usepackage[nocompress]{cite}
\else
  \usepackage{cite}
\fi
\ifCLASSINFOpdf
\else
\fi
\begin{document}

%
\title{Online POI Recommendation:  Learning Dynamic Geo-Human Interactions in Streams}
%
%
%
%

\author{Dongjie Wang,
        Kunpeng Liu,
        Hui Xiong,~\IEEEmembership{Fellow,~IEEE}
        Yanjie Fu,~\IEEEmembership{Member,~IEEE}
\IEEEcompsocitemizethanks{\IEEEcompsocthanksitem Dongjie Wang, Yanjie Fu are  with the Department
of Computer Science, University of Central Florida.
Emails: \{wangdongjie\}@knights.ucf.edu, yanjie.fu@ucf.edu.
\IEEEcompsocthanksitem Kunpeng Liu is with Department of Computer Science, Portland State University.
Email: kunpeng@pdx.edu
\IEEEcompsocthanksitem Hui Xiong is with Department of Management Science and Information Systems, Rutgers University.
Email: hxiong@rutgers.edu.
}
}

%
%

\markboth{Journal of \LaTeX\ Class Files,~Vol.~14, No.~8, August~2015}%
{Shell \MakeLowercase{\textit{et al.}}: Bare Demo of IEEEtran.cls for Computer Society Journals}
%



\IEEEtitleabstractindextext{%
\justifying
\begin{abstract}
Online POI recommendation strives to recommend future visitable places to users over time.
It is crucial to improve the user experience of location-based social networking applications (e.g., Google Maps, Yelp).
While numerous studies focus on capturing user visit preferences, geo-human interactions are ignored.
However, users make visit decisions based on the status of geospatial contexts.
In the meantime, such visits will change the status of geospatial contexts.
Therefore, disregarding such geo-human interactions in streams will result in inferior recommendation performance.
To fill this gap, in this paper, we propose a novel deep interactive reinforcement learning framework to model geo-human interactions.
Specifically, this framework has two main parts: the representation module and the imitation module.
The purpose of the representation module is to capture geo-human interactions and convert them into embedding vectors (state).
The imitation module is a reinforced agent whose job is to imitate user visit behavior by recommending next-visit POI (action) based on the state.
Imitation performance is regarded as a reward signal to optimize the whole interactive framework.
When the model converges, the imitation module can precisely perceive users and geospatial contexts to provide accurate POI recommendations.
Finally, we conduct extensive experiments to validate the superiority of our framework.
\end{abstract}

}

\maketitle


\IEEEdisplaynontitleabstractindextext

%
\IEEEpeerreviewmaketitle



 

\vspace{-0.3cm}
\IEEEraisesectionheading{\section{Introduction}\label{sec:introduction}}

Online Point of Interest (POI) recommendation aims to dynamically recommend next POIs to users over time by modeling human spatiotemporal activities at various POIs.
This kind of research can help users discover attractive places, provide personalized and user-centric human-technology interfaces (e.g., Google Maps, Foursquare, Yelp), improve user experiences and revenues in location-based social network applications, and, moreover, create a better geo-social community.

Most of the prior literature in POI recommendations takes a default setting: offline learning. With that being said, many offline-trained POI recommendation models~\cite{zhao2020go,chang2018content,hao2019real}  learn user visit preferences from historical check-in data and recommend next-visit POIs to users based on the static learned preferences.
However, the preferences and activity patterns of humans do indeed change over time. Online recommendation, in contrast to offline recommendation, assumes that user activity data (e.g., POI visits) are generated in real time, and that human preferences and learning environments change over time.
Moreover, by analyzing large-scale POI visit data, we observe that there is dynamic and mutual influence between users and geospatial contexts (e.g., POIs, POI categories, functional zones, etc.) in POI visit streams. For example, after visiting the New York City Metropolitan Museum, a user visits the Statue of Liberty in New York City. Each visit event will update the status of a user, as well as the status of the connected network of NYC tourism attractions. Users and the NYC tourism attraction network will mutually influence and be influenced over time, which we call geo-human interactions.
In the scenario of POI recommendation, modeling such varying and influential interactions in streams can help us to better understand how users evolve and how users influence and are influenced by a geospatial contextual and networked environment.
While there are studies in online recommendation~\cite{yang2020exploring,zou2020pseudo,chen2020efficient}, the integration of dynamic geo-human interactions and  online recommendation with streams is still in early stage.


To fill this gap, in this paper, we focus on modeling the geo-human interactions in streams for online POI recommendation.
We formulate the in-stream geo-human interaction modeling problem into a reinforcement learning task. 
The underlying idea is to regard a reinforcement learning agent as a recommender and regard the reinforcement learning environment as a joint composition of users and geospatial contexts (e.g., POIs, POIs categories, and functional zones), so that we can leverage the agent-environment interactions in reinforcement learning to model the interactions between users and geospatial contexts .  
To that end, we propose  a novel deep interactive reinforcement learning framework to unify both in-stream recommendation and geo-human interaction modeling.


Our proposed learning framework includes two major modules:  representation module and  imitation module. 
First, the representation module is to learn and track the representations of users and geospatial contexts. 
Second, the imitation module will use the representations of users and geospatial contexts as state and recommended POIs as actions to mimic how users visit places.
Third, the accuracy of the mimics of user visit patterns is fed back to the representation and imitation modules to help them optimize their parameters. 
When the imitation module perfectly mimics user visit patterns  the representation module produces the most accurate embeddings of interactive users and geospatial contexts.
Finally, the trained imitation module will choose the user-POI pairs with the highest Q value and make updated recommendations on the fly. 

Based on the overarching idea, we have conducted a preliminary study~\cite{DBLP:conf/aaai/WangW0ZHF21}.  
To simplify online recommendation, we only focused on modeling a single-user POI visit stream, instead of a mixed-user POI visit stream.
In particular, the representation module separately learned the representations of the user and the geospatial contexts.
Then, we concatenated the representations of a user and geospatial contexts together as the state of the environment.
The imitation module took the environment state as input and exploited its policy network to make a personalized next POI recommendation to the user.

However, the modeling of dynamic interaction in streams can be significantly improved.
Specifically, user visit intents exhibit \textit{multi-level interaction dynamics}:
1) \textit{Connected Dynamics}: a user's visit decision is impacted by other users and POI-related geographic entities through various types of visit events. Such a relationship forms a more comprehensive knowledge graph that includes not just spatial entities but also users.
2) \textit{Topological Dynamics}: new POI visit events continue to add new entities (users) and edges (visit events) to the knowledge graph.
3) \textit{Semantic Dynamics}: A user typically connects with POIs through various meta-path schemes based on distinct semantic relationships.
Therefore, it is appealing to model multi-level interaction dynamics for improving online POI recommendation in streams.

To this end, in our journal version, we consider a new data environment setup: a mixed-user visit stream. Based on the mixed-user visit stream, we construct a new dynamic knowledge graph, in which nodes are both users and geospatial entities, edges stand for user-user, geo-user, and geo-geo relationships, and, moreover, edges and nodes are added and deleted over time. This setup is fundamentally different from the data environment setup of our preliminary study: (1) a single-user visit stream; (2) nodes are only geospatial entities; (3) nodes and edges will not be deleted. 

However, the new data environment setup raises three algorithmic challenges. 
First, because there are millions of users, the new KG will grow exponentially larger over time. To address the exploding challenge, we  create an exit mechanism for long-ago visit-events in order to remove obsolete records for KG refinement.
Second, because the new KG will be larger and more sparse, it is challenging to identify POI candidates from a sparse and huge KG. 
To solve the problem of sparsity, we propose and use predefined meta-paths to choose the most relevant POI candidates for recommendation, taking into account semantics and the context of interests.
Finally, because the node set of the new KG is dynamic, it creates dynamic action space and makes recommendation policy difficult to train. 
To address this challenge, we develop a new deep neural policy network architecture to support the dynamic action space.

In summary, in this paper, we propose a deep interactive reinforcement learning framework for online POI recommendation. Our unique perspective is to model the multilevel dynamics of geo-human interactions in data streams. 
Specifically, our contributions are: 
(1) \textbf{Formulating the problem of modeling geo-human interactions in streams for online POI recommendation.}
We identify that human visit activities are an interactive, mutually-influenced process between users and geospatial entities.
We propose a new perspective to attack online POI recommendation by unifying in-stream recommendation and dynamic geo-human interaction modeling.

(2) \textbf{Proposing a representation-imitation based deep interactive reinforcement learning framework.}
To address the in-stream dynamic interaction modeling problem, we propose a novel deep interactive reinforcement learning framework. This framework has two main parts: the representation module and the imitation module. The former is to conduct dynamic knowledge graph learning to learn the state representations of users and geospatial entities. The latter is to perform online next-POI recommendations via policy networks. The two modules form a closed-loop interactive learning system.

(3) \textbf{Developing technical solutions to tackle multi-level interaction dynamics in mixed-user data streams.} 
We identify the difficulties that arise at three levels of dynamics in mixed-user data stream interactions: connected dynamics, topological dynamics, and semantic dynamics.
To address these challenges, we propose effective technical solutions. Firstly, we propose an exit mechanism to maintain the dynamic KG on a reasonable scale and reduce computational complexity.
Secondly, we develop a meta-path based candidate generation method for overcoming the challenge caused by sparse KG;
Thirdly, we devise a new policy network that generates Q-values for selecting recommended user-POI pairs to address the challenge of dynamic action space.

(4) \textbf{Extensive experiments with real world data.}
To validate the superiority of our framework, we conduct extensive experiments using two real-world datasets in comparison to six state-of-the-art baseline models. Experimental results show that our method consistently outperforms baselines to justify our technical insights.

\section{Definitions and Problem Statement}

\begin{figure*}[!t]
    \centering
    \includegraphics[width=0.7\linewidth]{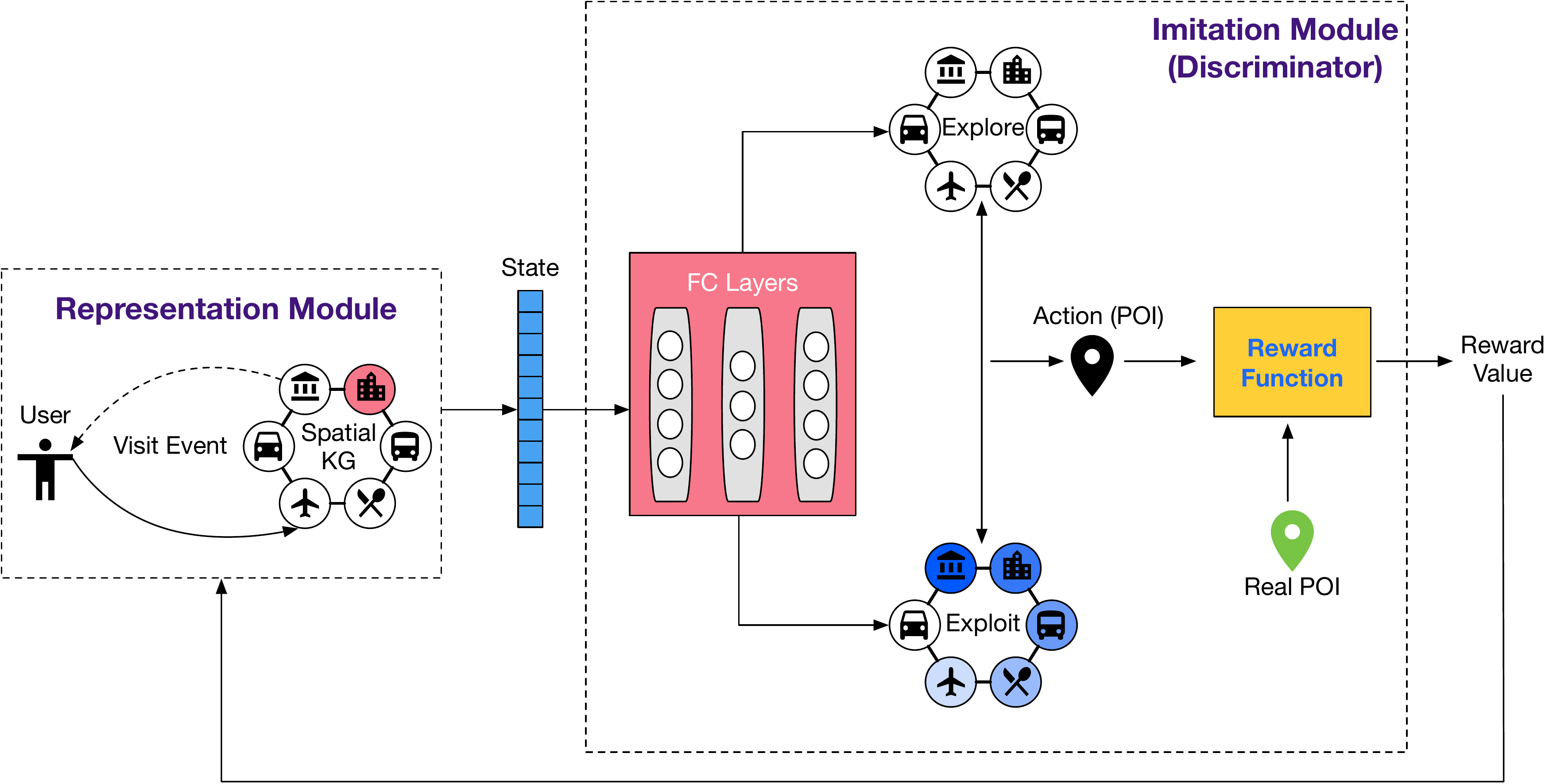}
    \captionsetup{justification=centering}
    \caption{The overview of the proposed framework. It consists of the representation module and the imitation module. The two modules mutually interact to capture geo-human interactions for recommending POIs accurately.}
    \vspace{-0.5cm}
    \label{fig:framework overview}
\end{figure*}

\vspace{-0.1cm}
\subsection{The Key Components of Our Framework}
We aim to address the joint task of  both online recommendation in streams and dynamic geo-human interaction modeling. We develop a deep interactive reinforcement learning framework that includes:

\begin{enumerate}
    \item {\bf Agent.} An agent is a recommender that predicts the next-visit POI of a given user, based on the current state of the environment (i.e., a knowledge graph of users and geospatial contexts).
    
    \item{\bf Action.} The action space refers to all possible visiting places (POIs). Let $a_i$ denotes the action at the $i$-th time step, assuming a user will visit the POI $P_i$ at the $i$-th time step, we expect $a_i = P_i$. 
    \item{\bf Environment.}
    The environment is where the recommendation of next-visit POIs is carried out. The joint and networked composition of users and geographic contexts (e.g., POIs, POI categories, functional zones, and their semantic linkages) is therefore regarded as the reinforcement learning environment. 
    We use a dynamic knowledge graph to represent the environment, with users, POIs, POI categories, and functional zones as nodes and visiting relations, affiliation relations, and locating relations as edges.

    \item{\bf State.} The state is a snapshot of the environment at a specific timestamp. 
    Formally, the dynamic KG at the step $l$ is denoted as $\mathcal{G}^l$.
    Then, the state $s^l$ refers to the representation vector of the $\mathcal{G}^l$.

    \item{\bf Reward.} The effectiveness of next visit POI recommendation is evaluated from three perspectives: POI-POI geographic distance, POI-POI category similarity,  and POI-POI identically equivalent indicator.
    We define the reward as the weighted sum of three parts:  (1) $r_d$, the reciprocal of the geographic distance between the real and predicted next-visit POI;  (2) $r_c$, the similarity between the real and predicted POI category.
    Here,  we utilize {\it GloVe}\footnote{\url{https://nlp.stanford.edu/projects/glove/}} embedding to calculate the cosine similarity between predicted and real POI categories as $r_c$; (3) $r_p$, the binary indicator shows whether the predicted visit POI is identically equivalent to the real one. In addition, to reduce the variance of the reward, we introduce the baselines ($b_d, b_c, b_p$) of the three elements into the reward. 
    Formally  we set a sliding window for $r_d$, $r_c$, $r_p$, respectively.
    The mean value of each sliding window is regarded as the baseline values of  $b_d, b_c, b_p$.
    Thus, the reward function can be formulated as follows: 
    \begin{equation}
        r = \sigma(\lambda_d \times (r_d-b_d) 
        + \lambda_c \times (r_c-b_c) + \lambda_p \times (r_p-b_p)),
    \end{equation}
    where $\lambda_d$ , $\lambda_c$ and $\lambda_p$ are weights for $r_d$, $r_c$ and $r_p$ respectively. 
    $\sigma$ is a sigmoid function, which aims to smooth  reward distribution for better policy learning.
\end{enumerate}

\subsection{Problem Statement}
We formulate the online POI recommendation problem into a joint task of recommendation in streams and dynamic geo-human interaciton modeling. 
We propose a deep interactive reinforcement learning framework.
In the framework, all users and geographic contexts (e.g., POIs, POI categories, functional zones) are considered as the environment, and a policy network is trained to replicate the decision-making process of user visits depending on the environment's state.
Formally, at the time step $l$, given the state of the environment $s^l$ , our propose is to find a mapping function $f : s^l \rightarrow a^{l+1}$.
Here, $a^{l+1}$ is the user next visit action in the time step $l+1$.
Thus, the function takes $s^l$ as input, and outputs $a^{l+1}$. 
During the mapping process, the state of the environment $s$ evolves over time. 
The objective is to model user-user, user-geo, geo-geo interactions and maximize the recommendation reward.

\section{Methodology}

\subsection{Framework Overview}
Figure \ref{fig:framework overview} shows our framework includes two modules: representation and imitation. 
First, the representation module is to learn the state of the environment.
In particular, it focuses on preserving the dynamic mutual interactions between user and geospatial contexts when learning the state embedding vector.
Second, the imitation module is to predict the next-visit POI and evaluate the reward value of this prediction.
In particular, it takes the state of the environment as input, and predicts the next-visit POI (action) of a given user.
The reward function is later exploited to evaluate the effectiveness of this recommendation.
After that, we leverage the reward value as a feedback signal to optimize the parameters of the representation and imitation modules.
The representation module and the imitation module enhance each other through iterative updates.
When the representation module learns the most accurate representation of the dynamic knowledge graph of users and geospatial contexts, the imitation module perfectly mimics users' next visit patterns.
Finally, we use the well-trained imitation module to perform online POI recommendations.

 \begin{figure*}[!t]
    \centering
    \includegraphics[width=0.85\linewidth]{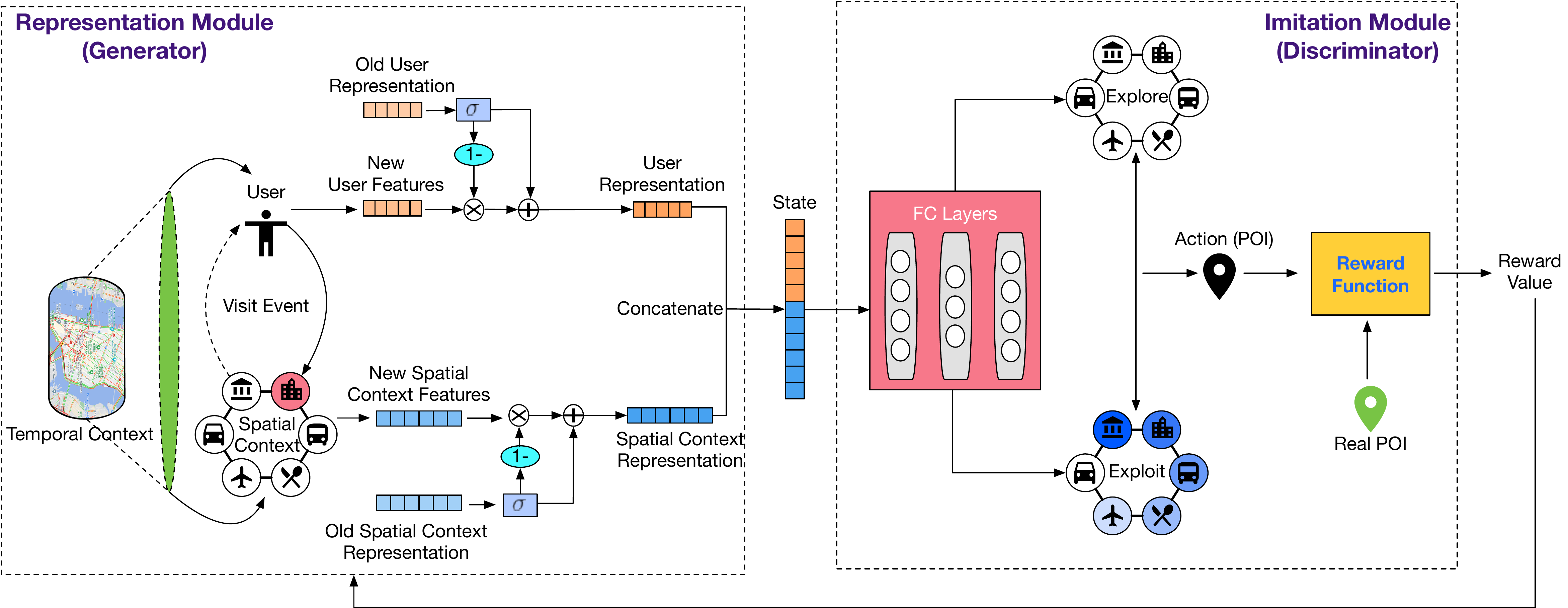}
    \captionsetup{justification=centering}
    \caption{The overview of the preliminary work. The representation module independently learns user and geospatial context embeddings. In the module, Geo-Human interactions are modeled using parameter updating rules. The imitation module recommends next-visit POI based on the state of the environment through exploration and exploitation.}
    \vspace{-0.5cm}
    \label{fig:aaai_framework}
\end{figure*}

\subsection{Static Spatial Knowledge Graph}
Our preliminary work~\cite{DBLP:conf/aaai/WangW0ZHF21} only considers geospatial contexts as a static spatial KG, learns the representations of users and geospatial contexts separately, and models the geo-human interactions exclusively in parameter updating rules.
Below, we first introduce the simplified method in our preliminary work. 

\subsubsection{Imitation Module}
Figure~\ref{fig:aaai_framework} shows the framework overview of our preliminary work.
The right part of this Figure shows that the imitation module takes the state of environment as input,  recommends next-visit POIs, and evaluates the recommendation effectiveness as reward. 
We use a classical Deep Q-Network (DQN)~\cite{mnih2013playing} as the backbone of the policy network. 
The DQN takes the state at the time $l$, denoted by $s^l$, as input and outputs the most possible visit POI based on $s^l$.
The $\epsilon-greedy$ method is used to integrate exploration and exploitation to capture user visit preferences sufficiently. 
In particular, the imitation module chooses a random POI $a^l_r$  with probability $\epsilon$, or selects the POI $a^l_m$ that owns the maximum $Q$ value with probability $1-\epsilon$, denoted by $a^l_m = \underset{a}{argmax}(Q(s^l,a^l))$. 
After that, the predicted POI and real POI are input into the reward function to calculate the reward value. Finally, the imitation module updates its parameters based on the Bellman Equation. The learning process continues until the imitation module fully imitates the users' visit behavior.

The imitation module has to search a large visiting POIs (action) space to learn user mobility patterns.
Thus, the learning process will be difficult and computationally expensive.
To alleviate this issue, we propose a training strategy to accelerate the learning procedure based on the prioritized experience replay~\cite{schaul2015prioritized}.
Specifically, we first assign a priority score for each data sample $(s^l,a^l,r^l,s^{l+1})$. Then, we construct a priority distribution based on the priority score for sampling batch of data from the memory.

We design two types of priority score: 
(1) \textbf{Reward-based}. 
In general, a larger reward value for a data sample indicates that it contributes more to policy learning.
Thus, we may set the reward value as the priority score of the data sample to improve learning.
The reward-based priority score at the $l$-th time step $x_r$ can be defined as 
\begin{equation}
    x_r = r^l 
\end{equation}

(2) \textbf{Temporal Distance (TD)-based}.
The TD error is originally used to update the policy network. 
A larger TD error for a data sample denotes that it contains more information not covered by the policy network.
Thus, we may set the TD error as the priority score of the data sample to make it can be sampled frequently for learning a better policy.
The TD-error based priority score at the $l$-th time step $X_{TD}$ can be defined as

\begin{equation}
    x_{TD}  =r^{l}+\gamma \max_{a^{l+1}}Q(s^{l+1}, a^{l+1}) - Q(s^{l}, a^{l})
\end{equation}

After assigning the priority score for each data sample, we calculate the distribution of the priority score.
Specifically, a softmax function takes the priority score as input and outputs the priority distribution.
Then, we sample top $K$ data samples based on the distribution as one batch to train the imitation module.

\begin{figure*}[!t]
    \centering
    \includegraphics[width=0.85\linewidth]{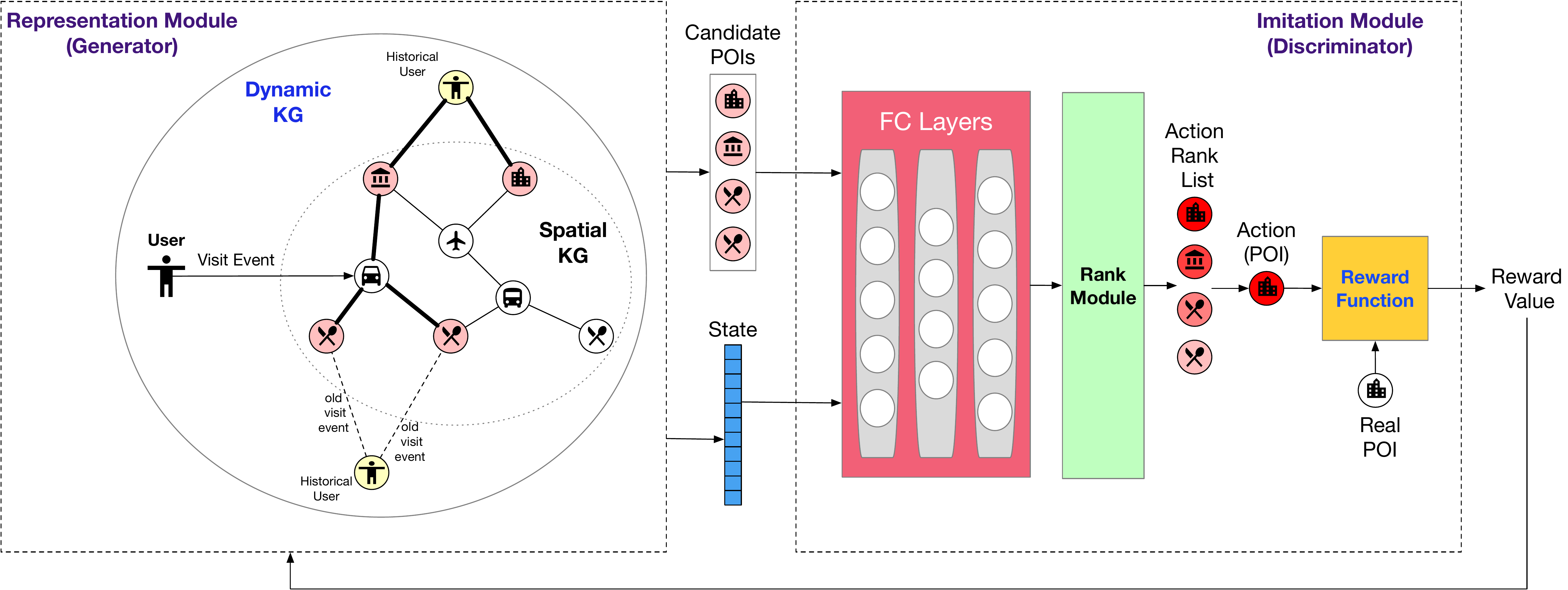}
    \captionsetup{justification=centering}
    \caption{The overview of the work in the journal version. A dynamic KG is utilized to integrate users, geospatial contexts, and geo-human interactions. The representation module learns the embedding of the dynamic KG as the state. Candidate POIs are selected to capture mobility semantics and improve recommendation performance. The state and candidate POIs are given to the imitation module, and the next-visit POI is given back. }
    \vspace{-0.5cm}
    \label{fig:tbd_framework}
\end{figure*}

\subsubsection{Representation Module}
The left part of Figure \ref{fig:aaai_framework} shows that the representation module considers two types of information 
(1) the interactions between users and geospatial contexts;
(2) the temporal dependency of user representations; 
in the updating rules.

First,  the spatial knowledge graph ({\it KG}) indicates the geospatial contexts.
We denote the spatial {\it KG} as
$\mathbf{g}^l = < \mathbf{h}^l, \mathbf{rel}, \mathbf{t}^l >$,
where $\mathbf{h}^l$ is the representation of the heads ({\it i.e.}, POIs), $\mathbf{t}^l$ is the representation of the tails ({\it i.e.}, categories and functional zones),
$\mathbf{rel}$ is the representation of the relationship between the heads and tails.
Second, the temporal context $\mathbf{T} \in \mathbb{R}^{M \times 3}$ is the combination of inner traffic, in-flow traffic, and out-flow traffic in all areas of a city, where  $M$ represents the number of areas, and $3$ represents the three kinds of traffic flows.
The interactions between users and geospatial \textit{KG} occur concurrently with the temporal context.
The users' visit events change the representation of spatial {\it KG}, and a new spatial {\it KG} representation leads users to choose next-visit places.
Meanwhile, new user representations are not just related to current user visit preference changes but also associated with the representative parts of  old user visit interests.

{\bf Updating Rules of User Representations} Assuming a user $u_i$ visits the POI $P_j$ at the step $l$, the user representation will be updated for the step $l+1$. We incorporate the user representation $\mathbf{u}_i^l$ and the interactions between the POI $P_j$ and the user $u_i$ into $\mathbf{u}_i^{l+1}$ such that:
$
    \mathbf{u}_i^{l+1} = \sigma(\alpha_u \times \mathbf{u}_i^{l} + 
    (1-\alpha_u) \times (\mathbf{W}_u \cdot (\mathbf{h}_{P_j}^{l})^{\intercal} \cdot  \tilde{\mathbf{T}}^l)),
$
where
$\mathbf{W}_u \in \mathbb{R}^{N \times 1}$ is the weight; $\alpha_u$ is a scalar that represents the proportion of old profiling information in $\mathbf{u}_i^{l+1}$ , it is given by:
$
    \alpha_u = \sigma(\mathbf{W}_{\alpha_u} \cdot \mathbf{u}_i^l + \mathbf{b}_{\alpha_u}),
$
where $\mathbf{W}_{\alpha_u} \in \mathbb{R}^{1 \times N}$ is the weight and $\mathbf{b}_{\alpha_u} \in \mathbb{R}^{1 \times 1}$ is the bias term; 
$\tilde{\mathbf{T}}^{l} \in \mathbb{R}^{N\times1}$ is the temporal context vector adaptable with state update, it can be calculated by:
$
    \tilde{\mathbf{T}}^l = \sigma(\mathbf{W}_{T_1} \cdot \mathbf{T}^{l} \cdot \mathbf{W}_{T_2} + \mathbf{b}_T),
$
where  $\mathbf{W}_{T_1} \in \mathbb{R}^{N\times M}$, $\mathbf{W}_{T_2} \in \mathbb{R}^{3\times1}$ and $\mathbf{b}_T \in \mathbb{R}^{N\times1}$ are the weights and bias respectively.

{\bf Updating Rules of Spatial KG Representations}.
During the learning process of the representation of the spatial {\it KG}, we only focus on directly visited POI $P_j$ and other POIs $P_{j^-}$ that ``belong to" the same category or ``locate at" the same functional zones with the directly visited $P_j$.
Here, heads are POIs and tails are categories or functional zones.
Formally, we need to update the spatial {\it KG} representation $\mathbf{g}^l = < \mathbf{h}^l, \mathbf{rel}, \mathbf{t}^l >$.
We update the information in  $\mathbf{h}^l$ and $\mathbf{t}^l$, except  $\mathbf{rel}^l$.
In addition, $\sigma{(\cdot)}$ denotes the sigmoid function in following formulas.

\begin{enumerate}[(1)]
    \item Updating visited POI $\mathbf{h}_{P_j}^{l+1}$. Similar to update  $\mathbf{u}^{l+1}_i$, we incorporate the old visited POI representation $\mathbf{h}^{l}_{P_j}$ and the interactions between the use $\mathbf{u}_i^l$ and the POI $P_j$ in a given temporal context:
    $
           \mathbf{h}_{P_j}^{l+1} = \sigma(\alpha_p \times \mathbf{h}_{P_j}^{l} + (1-\alpha_p) \times (\mathbf{W}_p \cdot (\mathbf{u}_{i}^{l})^{\intercal} \cdot  \tilde{\mathbf{T}}^l)),
    $
    where $\mathbf{W}_p \in \mathbb{R}^{N \times 1}$ is the weight;
    $\alpha_p$ is a scalar that denotes the proportion of old POI information in $\mathbf{h}_{P_j}^{l+1}$, it is calculated by:
    $
        \alpha_p = \sigma(\mathbf{W}_{\alpha_p} \cdot \mathbf{h}_{P_j}^l +\mathbf{b}_{\alpha_p}),
    $
    where $\mathbf{W}_{\alpha_p} \in \mathbb{R}^{1 \times N}$ is weight and $\mathbf{b}_{\alpha_p} \in \mathbb{R}^{1 \times 1}$ is bias.
    

    \item Updating category and functional zones (tail) $\mathbf{t}_{P_j}^{l+1}$. We update tail $\mathbf{t}_{P_j}^{l+1}$ by the combination of $\mathbf{t}_{P_j}^{l}$, $\mathbf{h}_{P_j}^{l+1}$ and $\mathbf{rel}_{P_j}$, it can be denoted as:
    $
    \mathbf{t}_{P_j}^{l+1} =
    \alpha_t \times \mathbf{t}_{P_j}^{l} +
    (1-\alpha_t) \times (\mathbf{h}_{P_j}^{l+1} + \mathbf{rel}_{P_j}).
   $
    where $\alpha_t$ is scalar that denotes the proportion of old tail information in $\mathbf{t}_{P_j}^{l+1}$, it is calculated by:
    $
    \alpha_t = \sigma(\mathbf{W}_{\alpha_t} \cdot \mathbf{t}_{P_j}^{l} +\mathbf{b}_{\alpha_t}),
    $
    where $\mathbf{W}_{\alpha_t} \in \mathbb{R}^{1 \times N}$ is weight and $\mathbf{b}_{\alpha_t} \in \mathbb{R}^{1 \times 1}$ is bias. 
    
    \item Updating same category and location POIs $\mathbf{h}^{l+1}_{P_{j-}}$
    
    We update the other POIs that belong to the same POI category and locate at the same functional zones of the visited $P_j$ as
    \small
    \begin{equation}
    \left\{
             \begin{array}{lr}
             \mathbf{h}^{h}_{P_{j^-}} =\mathbf{t}_{P_{j^-}}^{l+1} -  \mathbf{rel}_{P_{j^-}}, &  \\
         \mathbf{h}^{l+1}_{P_{j^-}}=
             \sigma[\alpha_{h} \times \mathbf{h}^{l}_{P_{j^-}} + (1-\alpha_{h}) \times \mathbf{h}^{h}_{P_{j^-}} ]
             \end{array}
    \right.
    \end{equation}
    where $\alpha_h$ is a scalar that is the proportion of  $\mathbf{h}^{l}_{P_{j^-}}$ in $\mathbf{h}^{l+1}_{P_{j^-}}$, it is calculated by:
    $
    \alpha_h = \sigma(\mathbf{W}_{\alpha_h} \cdot \mathbf{h}_{P_{j^-}}^{l} +\mathbf{b}_{\alpha_h}),
    $
    where $\mathbf{W}_{\alpha_h} \in \mathbb{R}^{1 \times N}$ is weight and $\mathbf{b}_{\alpha_h} \in \mathbb{R}^{1\times 1}$ is bias.
    After the above updating and learning process, we concatenate $\mathbf{u}_i^{l+1}$ and $\mathbf{g}^{l+1}$ as the state $s^{l+1}$. 
\end{enumerate}

\subsection{Dynamic Geo-Human Knowledge Graph}

\subsubsection{Advancing Representation Module}

Figure \ref{fig:tbd_framework} shows that, we propose an enhanced representation module that unifies both users and geospatial contexts (POIs, POI categories, functional zones) into a single dynamic attributed knowledge graph, in order to better model connected, topological, and semantic dynamics. 

\noindent\textit{\underline{Constructing A New Dynamic KG.}}
The dynamic KG construction includes two stages: 1) \textbf{initialization stage} and 2) \textbf{evolving stage}.
In the initialization stage, we construct a classical knowledge graph to depict the semantic connectivity among different spatial entities.
There are three types of spatial entities in the stage: POI, POI category, POI location ({\it i.e. functional zones}), and two relations types: ``belong to'' and ``locate at''.
We organize two triple facts based on the entities and relations: (1) $<$POI, "belong to", POI category$>$, which expresses the affiliation relation between POI and POI category; (2) $<$POI, "locate at", functional zone$>$, which demonstrates the geographical  relation between POI and function zone.
In the evolving stage, we aim to add the dynamic interactions (edges) between user and spatial entities.
We introduce two entities: user, RPOI, and two relations: ``visit'', ``also visit'' into the KG that is constructed in the initialization stage. 
Here, RPOI is the reduplicated version of POI for recording the ``also visit'' relation.
When a user visit event occurs, we add two triple facts to the KG: (1) $<$user, "visit", POI$>$, which shows the visit relation between the user and POI; (2) $<$POI, "also visit", RPOI$>$, which indicates the visit cascade relation between different POIs. 
In addition, to reflect the temporal effect of visit events, we record the visit time of each ``visit'' relation.
To show the popularity of POIs, we record the total visits of each POI.

\noindent\textit{\underline{Exit Mechanism for Expired Information.}}
Since millions of users and POIs are added to the KG over time, the size of the KG will become uncontrollable, resulting in memory overflow and huge computational complexity.
We found that latest user activities are more significant to depict user visit preferences in comparison to old user trajectories.
Thus, we implement an exit mechanism
to remove expired users and redundant activity information. 
Specifically, for each user, we create a sliding window with a fixed length to record visit events.
When a new user visit event occurs, if the user's number of visit events does not exceed the window size, the new event is added to the window; otherwise, expired information is removed from the window before the new event is added.
The benefits of this idea are that the dynamic KG can keep a reasonable size and retain the most recent user visit preferences as the time step increases.

\noindent\textit{\underline{Learning Representations of Dynamic KG.}}
Since traditional policy networks take vectors/matrices as input, we need to transform  the environment (dynamic KG) into a state representation vector. 
Traditional KG embedding methods are not suitable for dynamic KG, because they have to retrain from scratch when new entities or relations are added to the dynamic KG.

\begin{figure}[t]
	\centering\includegraphics[width=0.86\linewidth]{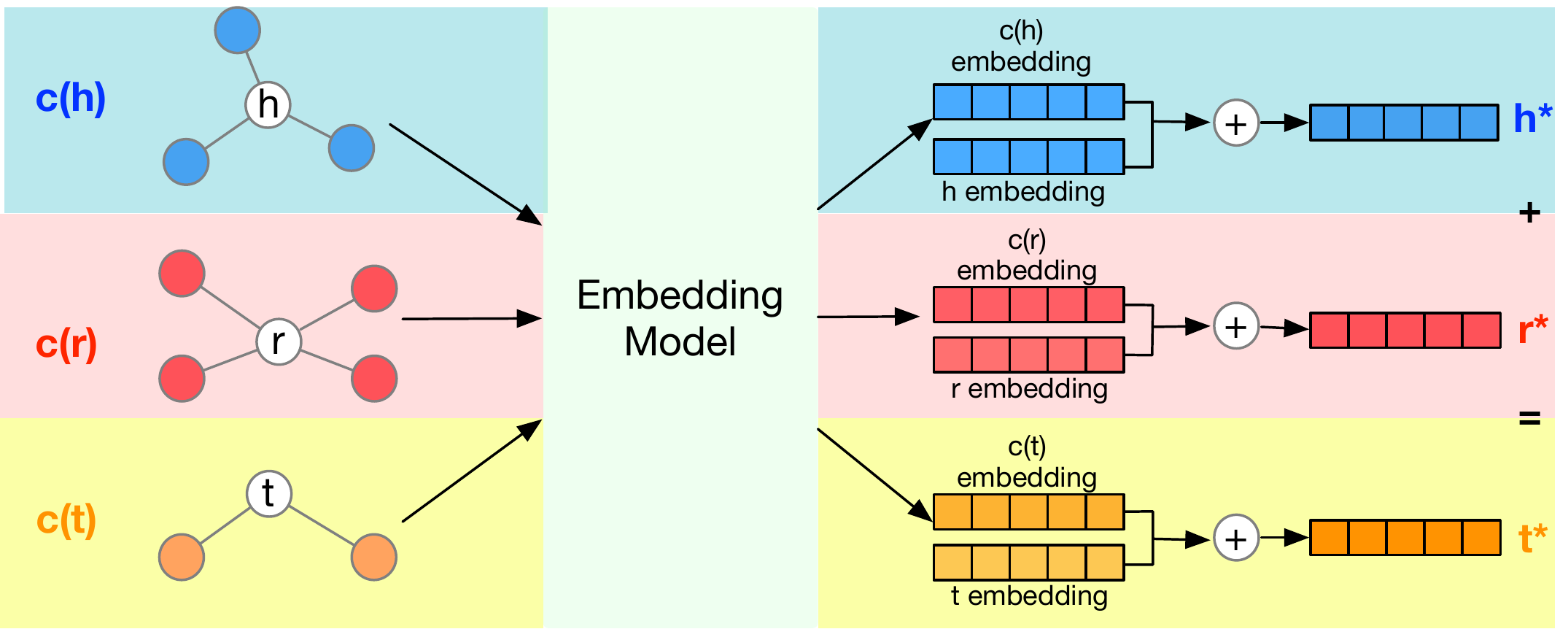}
	\caption{Representation learning for dynamic KG.}
	\label{fig:dkgr}
	\vspace{-0.5cm}
\end{figure}

To overcome this limitation, we leverage the technique in ~\cite{wu2019efficiently}. Figure \ref{fig:dkgr} shows that we obtain the embedding of the dynamic KG by preserving the translation relationship $\mathbf{h}^* + \mathbf{r}^* = \mathbf{t}^*$, where $\mathbf{h}^*, \mathbf{r}^*, \mathbf{t}^*$ are the joint embedding of head entity, relation, and tail entity respectively. 
The joint embedding of object(i.e. entities and relations) in the dynamic KG is the combination of the embedding of itself (i.e. \textbf{h}, \textbf{r}, \textbf{t}) and its context (i.e. $\mathbf{cx}(h), \mathbf{cx}(r), \mathbf{cx}(t)$.
The context of an entity includes itself and its one-hop neighbor entities.
The context of a relation consists of itself and other relations connecting the same entity pairs.

Be sure to notice that, the representation learning is performed concurrently with the construction of the dynamic KG.  
In the initialization stage of dynamic KG construction, we aim to preserve the translation relationship $\mathbf{h}^* + \mathbf{r}^* = \mathbf{t}^*$ among the joint embeddings of all objects (i.e entity and relation) in the KG.
Let $o$ be an entity, the context of $o$ is a sub-graph structure in the KG.
We first input the context of $o$ into graph convolutional networks (GCNs) with $m$ layers.
The embedding of the final layer can be represented as
\begin{equation}
    \mathbf{Z}^{m} = \text{relu} ( \hat{\mathbf{D}}^{-\frac{1}{2}} \hat{\mathbf{A}} \hat{\mathbf{D}}^{-\frac{1}{2}} \mathbf{Z}^{m-1} \mathbf{W}^{m-1}),
\end{equation}
where relu is the activation function, $\hat{\mathbf{A}}=\mathbf{A}+\mathbf{I}$, $\mathbf{A}$ is the adjacency matrix, $\mathbf{I}$ is the identity matrix, $\hat{\mathbf{D}}$ is the degree matrix, and $\mathbf{W}^{m-1}$ is the weight matrix.
For the learned embedding
 $\mathbf{Z}^m \in \mathbb{R}^{n \times d}$,  $n$ is the number of vertex in the sub-graph and $d$ is the dimension of each row in $\mathbf{Z}^m$.
Then, we employ an attention layer to aggregate   $\mathbf{Z}^m$ into a graph-level embedding.
The attention weight for each vertex in the sub-graph can be calculated by:
\begin{equation}
    \alpha_i = \frac{\text{exp}(\text{score}(\mathbf{Z}^m_i,\mathbf{o}))}{\sum_{i=1}^n \text{exp}(\text{score}(\mathbf{Z}^m_i,\mathbf{o}))}
\end{equation}
where $\mathbf{Z}_i^m$ is the embedding of the $i$-th vertex in the sub-graph, $\mathbf{o} \in \mathbb{R}^d$ is the embedding of the object,
$\text{score}(.)$ measures the relevance between $\mathbf{Z}^m_i$ and $\mathbf{o}$. 
Next, we obtain the context embedding $\mathbf{cx}(o)$ by calculating the weighted sum of all vertex embeddings, which can be denoted as:
\begin{equation}
    \mathbf{cx}(o) = \sum_i^n \alpha_i \mathbf{Z}_i^m
\end{equation}
Later, we incorporate $\mathbf{o}$ with $\mathbf{cx}(o)$ as the joint embedding $\mathbf{o}^*$   through a gated neural cell as follows:
\begin{equation}
    \mathbf{o}^* = \sigma(\bm{\gamma}) \odot \mathbf{o} + (1-\sigma(\bm{\gamma})) \odot \mathbf{cx}(o)
\end{equation}
where $\sigma$ is the sigmoid function, $\odot$ is element-wise multiplication, $\bm{\gamma}$ is a trainable parameter vector.
We minimize the following loss function during the training process:
\begin{equation}
    \mathcal{L} = \sum_{(h,r,t)\in S} \sum_{(h',r,s')\in S'} max(0,f(h,r,t)+ \varepsilon - f(h',r,t'))
\end{equation}
where $f(h,r,t) = \left \| \mathbf{h}^* + \mathbf{r}^* - \mathbf{t}^* \right \|_{l1}$, $\varepsilon$ is a tolerance margin value, $S$ is the positive triple set that contains the correct semantic meaning, $S'$ is the negative triple set that randomly replaces the head and tail entities in $S$. 
After that, we apply the average pooling on both entity and relation embeddings to obtain the representation of the whole KG.
We regard the representation as the initial state of the environment for the reinforced agent of the imitation module, denoted by $\mathbf{s}^0$.

In the evolving stage of dynamic KG construction, we learn the KG embedding by partially updating  $\mathbf{s}^0$ according to the information of new events.
In particular,  we incrementally learn the dynamic KG embedding via a local updating strategy.
The initial KG embedding has been obtained by the initialization stage.
As new nodes are added or old nodes are removed, only the translation relationships of impacted items will be broken.
Other items whose contexts are unaffected by the event still maintain original translation relations.
Therefore, we pick up all objects that are affected by the node update event and utilize the same model structure that is used in the initialization stage to learn the embedding of these objects incrementally.
In this way, we avoid retraining the whole model from the scratch.
The state evolves over time: $\mathbf{s}^0 \rightarrow \mathbf{s}^1 \rightarrow \cdots \rightarrow \mathbf{s}^T$.

\subsubsection{Advancing Imitation Module}

\noindent\textit{\underline{POI Candidates Generated by Meta-path.}}
Intuitively, different users have different mobility patterns and visiting preferences.
They decide where to go on their next trip for a variety of reasons, such as proximity, recommendations from friends, and how the POI works.
So, we can generate POI candidates based on such prior knowledge to reduce the action space and improve the recommendation performance.
To achieve this goal, we propose a meta-path based POI candidate generation method.


Here, we define four kinds of  meta-path scheme in the dynamic KG:
(1)  ``user --$>$ visit --$>$ POI''; (2)  ``user --$>$visit --$>$ POI --$>$ also visit --$>$ RPOI''; (3)  ``user --$>$ visit --$>$ POI --$>$ belong to --$>$POI category --$>$ belong to --$>$POI''; (4) ``user --$>$ visit --$>$POI --$>$ locate at --$>$ functional zone --$>$ locate at --$>$ POI''.
These paths all start from the ``user'' entity and end with the ``POI'' entity.
To simplify the discussion, we take the POI recommendation at the $l$-th time step as an example.
For a given user, we generate POI candidates by applying the four meta-paths on the dynamic KG.
Each meta-path produces a POI set according to the corresponding schema.
Then, the Top-K popular POIs of each POI set are collected as the POI candidates at the time step $l$, denoted by $cand^l \in \mathbb{R}^{4K}$, where $K$ represents the number of POI candidates of each predefined meta-path.
In this way, we reduce the action space from all POIs to $4K$. 

\begin{figure}[t]
	\centering\includegraphics[width=0.7\linewidth]{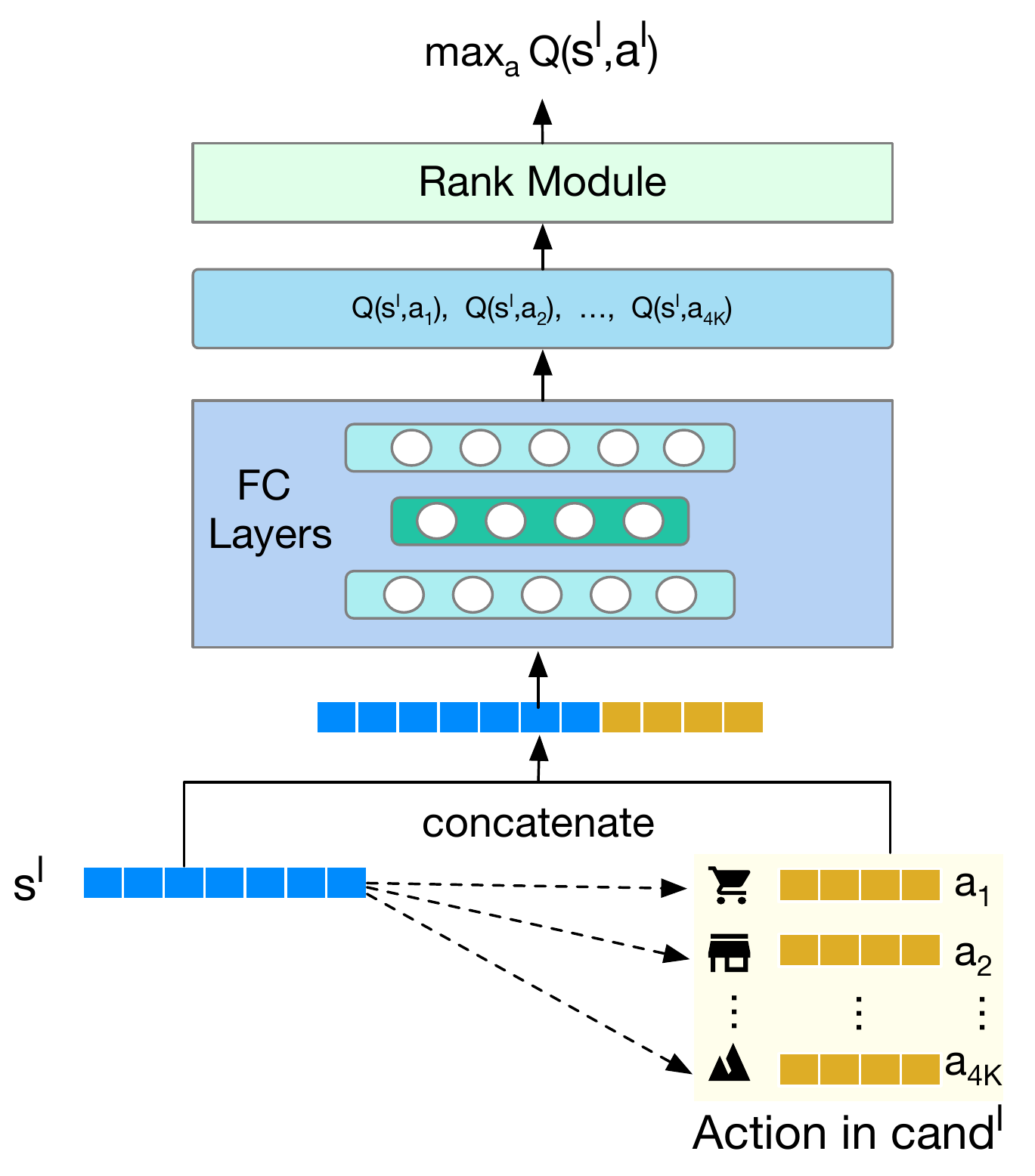}
	\vspace{-0.2cm}
	\caption{The structure of the enhanced policy network. The network learns the Q-value of the concatenation of the state and action embedding. The action with the highest Q-value will be regarded as the next-visit POI.}
	\label{fig:DQN}
	\vspace{-15pt}
\end{figure}

\noindent\textit{\underline{A New Policy Network Structure for Dynamic Action Space.}}
In our preliminary study, we exploit a vanilla DQN as a policy network
However, the vanilla DQN cannot deal with a dynamically varying action space, because it only learns a fixed point-wise mapping from a state to an action in a fixed action set.
In the journal version, the action space will change over time since different users have different POI candidates.
To overcome this challenge, we propose a new policy network structure as shown in Figure \ref{fig:DQN}.

The underlying idea of this  network structure is to learn a pairwise mapping function that maps a state-action embedding pair to a score. 
To simplify the description, we 
take the $l$-th time step as an example.
Specifically, we first obtain the state embedding $\mathbf{s}^l$ and the POI candidates ${cand}^l=\{ \mathbf{a}_1, \mathbf{a}_2, ..., \mathbf{a}_{4K} \}$.
Then, we concatenate the state embedding with each action embedding in ${cand}^l$ respectively, then regard the concatenation as the input of the fully connected (FC) layers.
The outputs of the FC layers are   the Q-values of all state-action pairs in ${cand}^l$.
Finally, we feed all Q-values into a rank module to find the action with the highest Q-value and regard it as the  action $a^l$.

\subsection{Solving the Optimization Problem}
Our method is a closed-loop learning system. 
We propose an adversarial training-like optimization method to train the model sufficiently.  
We interpret our method from an adversarial learning perspective.
We regard the representation module as a generator to produce the state in real-time.
Then, we treat the imitation module and reward function as a discriminator.
When the discriminator always gives the highest score for quality, the best situation for the representation module is reached.
As opposed to classical adversarial learning, the scoring criteria of our method are deterministic.
The reward value of each imitation behavior indicates the imitated performance.
Thus, we regard the reward value as feedback to update the parameters of the representation module for improving representation learning capability.
We utilize the gradient that comes from the gap between the current reward and the expected reward value to update the parameters of the representation module.


  
    

\section{Experiments}
We conducted experiments on two real-world datasets to answer the following questions:  \textbf{Q1.} Does our work outperform the existing methods?  \textbf{Q2.} How about the robustness of the proposed framework?  \textbf{Q3.} Is each part of the proposed framework effective for improving recommendation performance? \textbf{Q4.} How does the reward function impact the POI recommendation performance?

\subsection{Experimental Setup}
\subsubsection{Data Description}
Table \ref{table:data_stat} shows the statistics of two check-in datasets: New York and Tokyo \cite{yang2014modeling}.
Each dataset includes User ID, Venue ID, Venue Category ID, Venue Category Name, Latitude, Longitude, and Timestamp. 
\begin{table}[t!hbp]
	\centering
	\scriptsize
	\tabcolsep 0.04in
	\caption {Statistics of the checkin data.}
	\begin{tabular}[t]{c|c|c|c|c}
		\hline
		\textbf{City}         & \textbf{\# Check-ins} & \textbf{\# POIs} &\textbf{\# POI Categories} & \textbf{Time Period}  \\ \hline
		New York & $227,428$ & $38,334$ &251 & 4/2012-2/2013 \\ \hline
		Tokyo & $537,703$ & $61,858$ &385 & 4/2012-2/2013 \\
		\hline
	\end{tabular}
	\label{table:data_stat}
\end{table}

\begin{figure*}[!thb]
	\centering
	\subfigure[Precision on Category]{\label{fig:overall_prec_cat_nyc}\includegraphics[width=4.35cm]{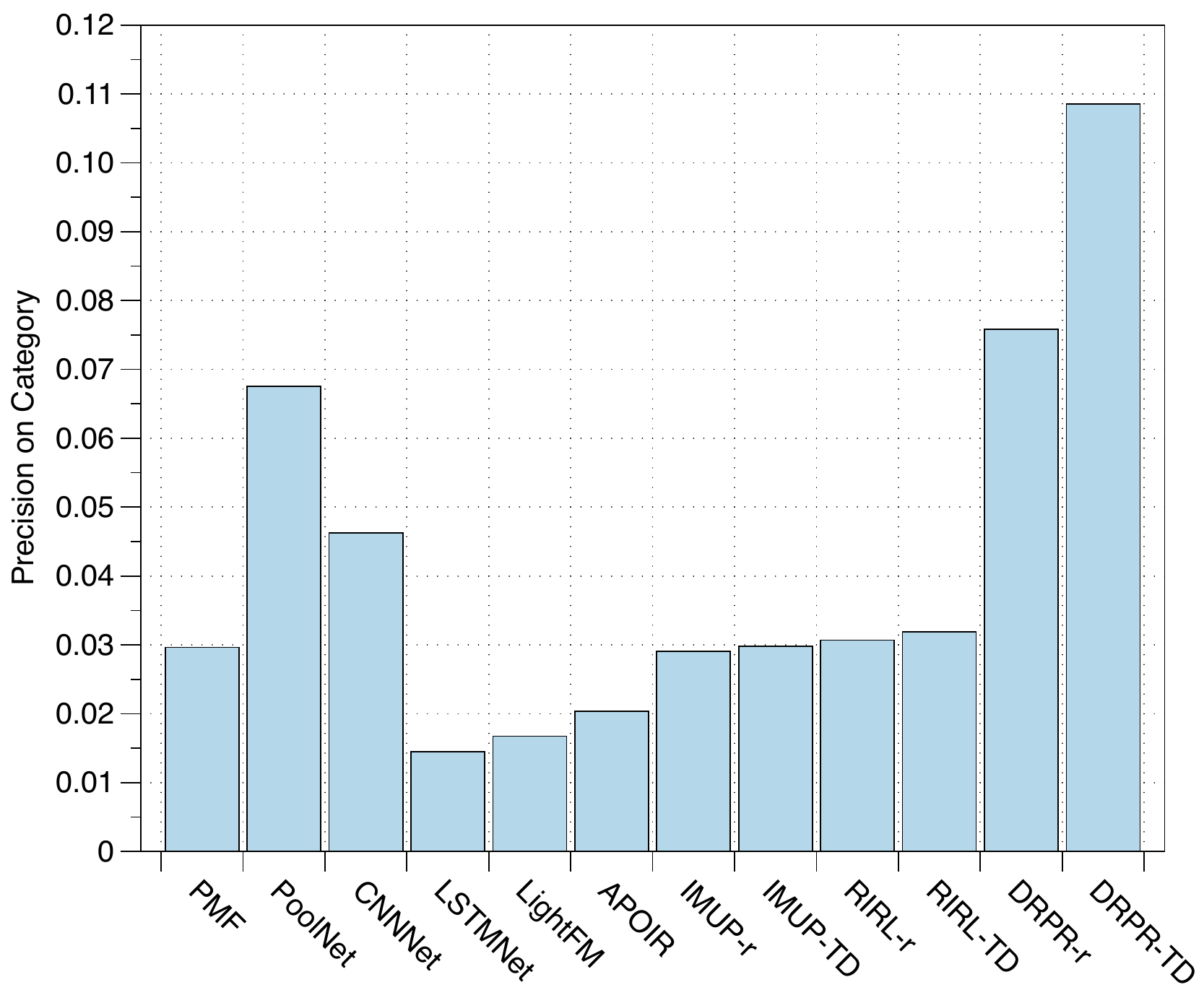}}
	\subfigure[Recall on Category]{\label{fig:overall_rec_cat_nyc}\includegraphics[width=4.35cm]{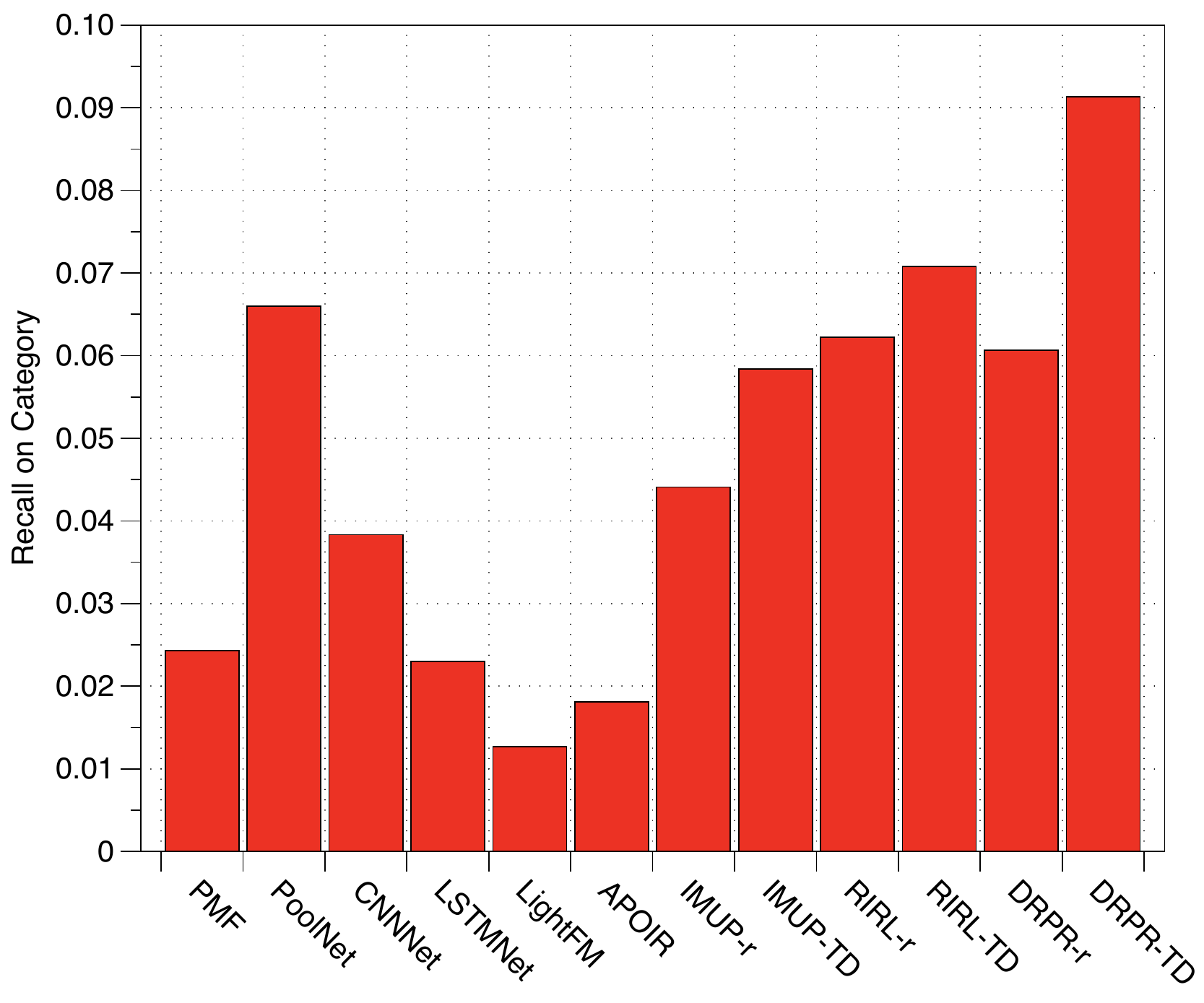}}
	\subfigure[Average Similarity]{\label{fig:overall_avg_sim_nyc}\includegraphics[width=4.35cm]{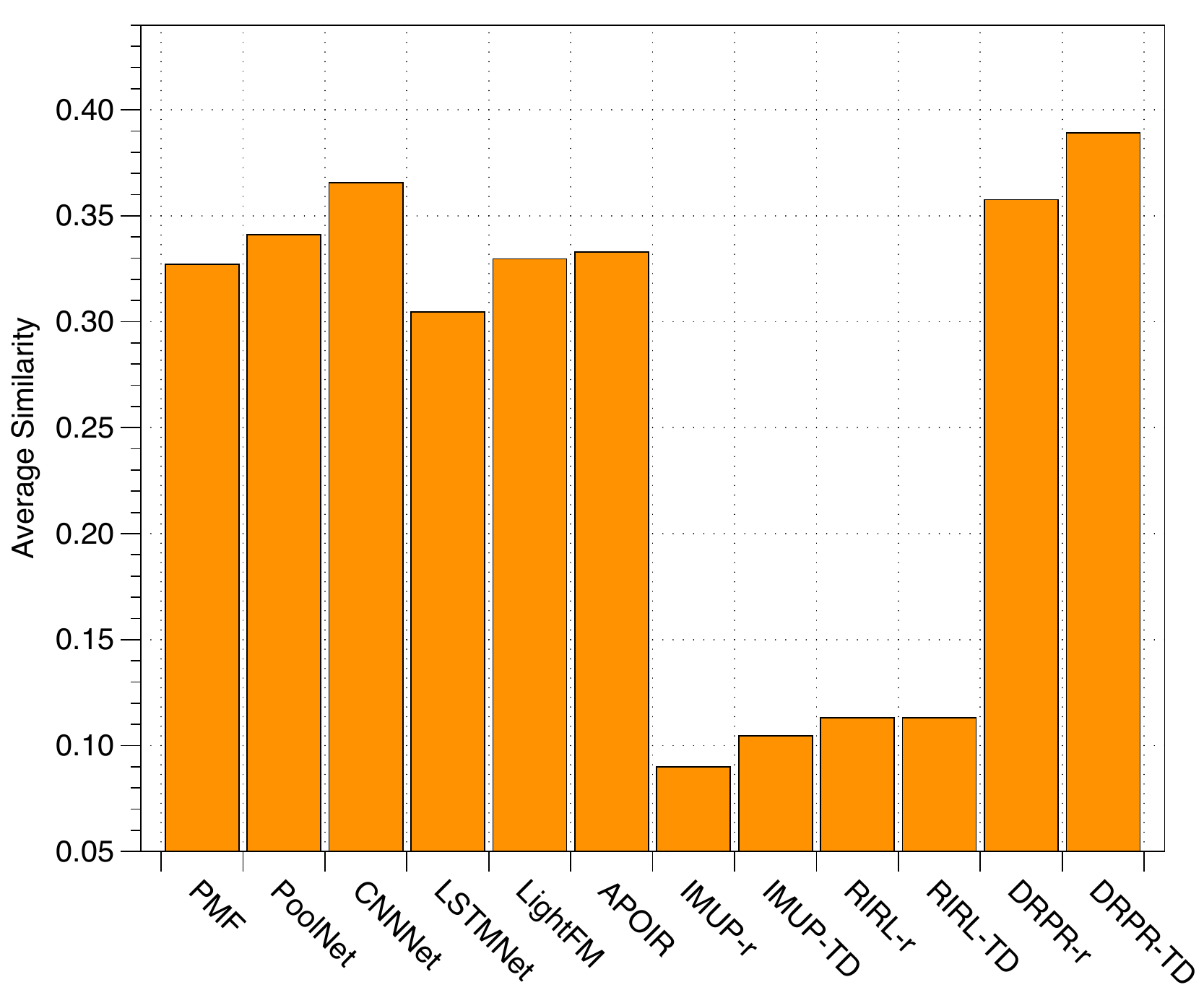}}
	\subfigure[Average Distance]{\label{fig:overall_avg_dis_nyc}\includegraphics[width=4.35cm]{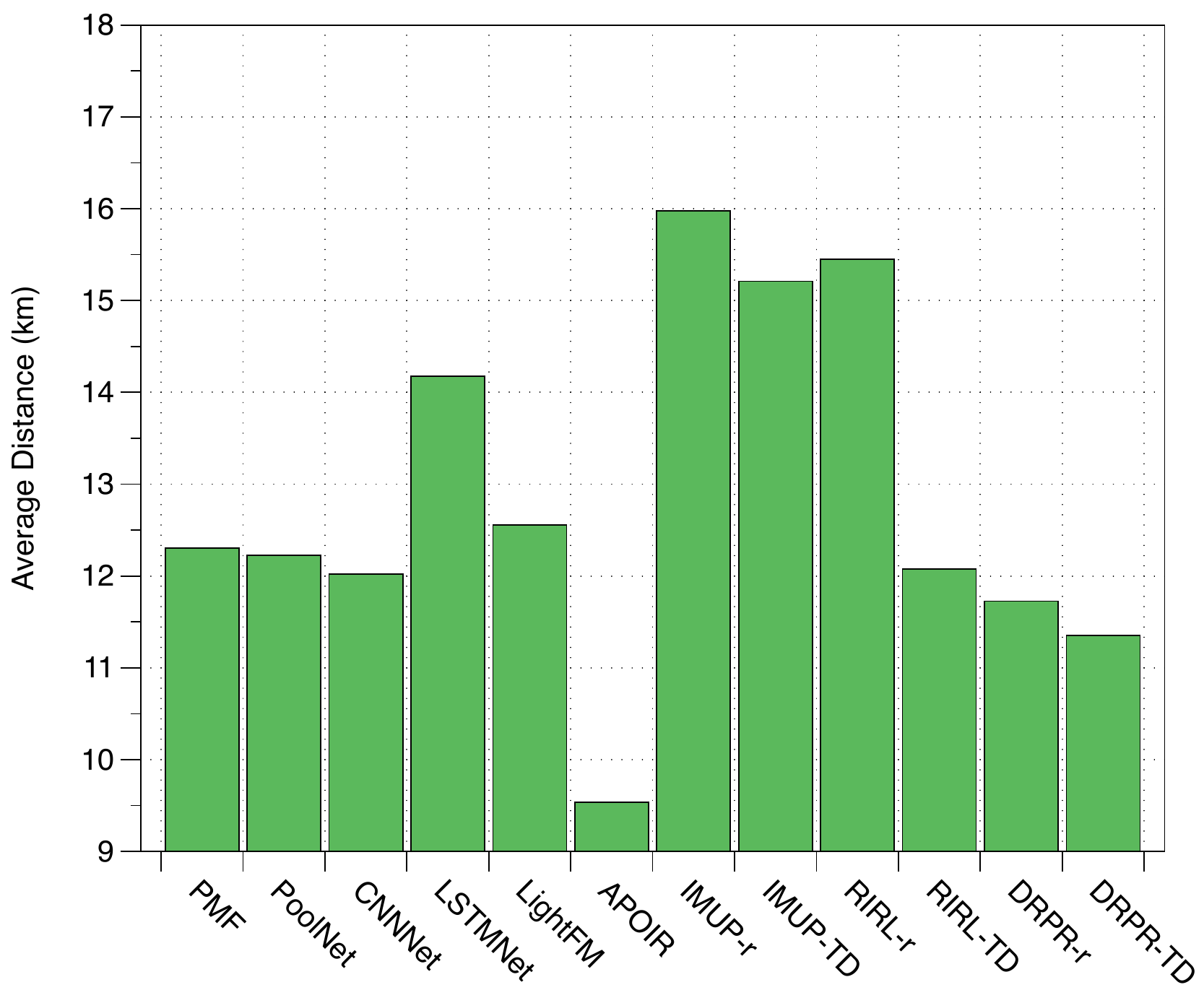}}
		\vspace{-0.3cm}
	\captionsetup{justification=centering}
	\caption{Overall performance {\it w.r.t.} New York dataset.}
		\vspace{-0.5cm}
	\label{fig:nyc_overall}
\end{figure*}

\begin{figure*}[!thb]
	\centering
	\subfigure[Precision on Category]{\label{fig:overall_prec_cat_tky}\includegraphics[width=4.35cm]{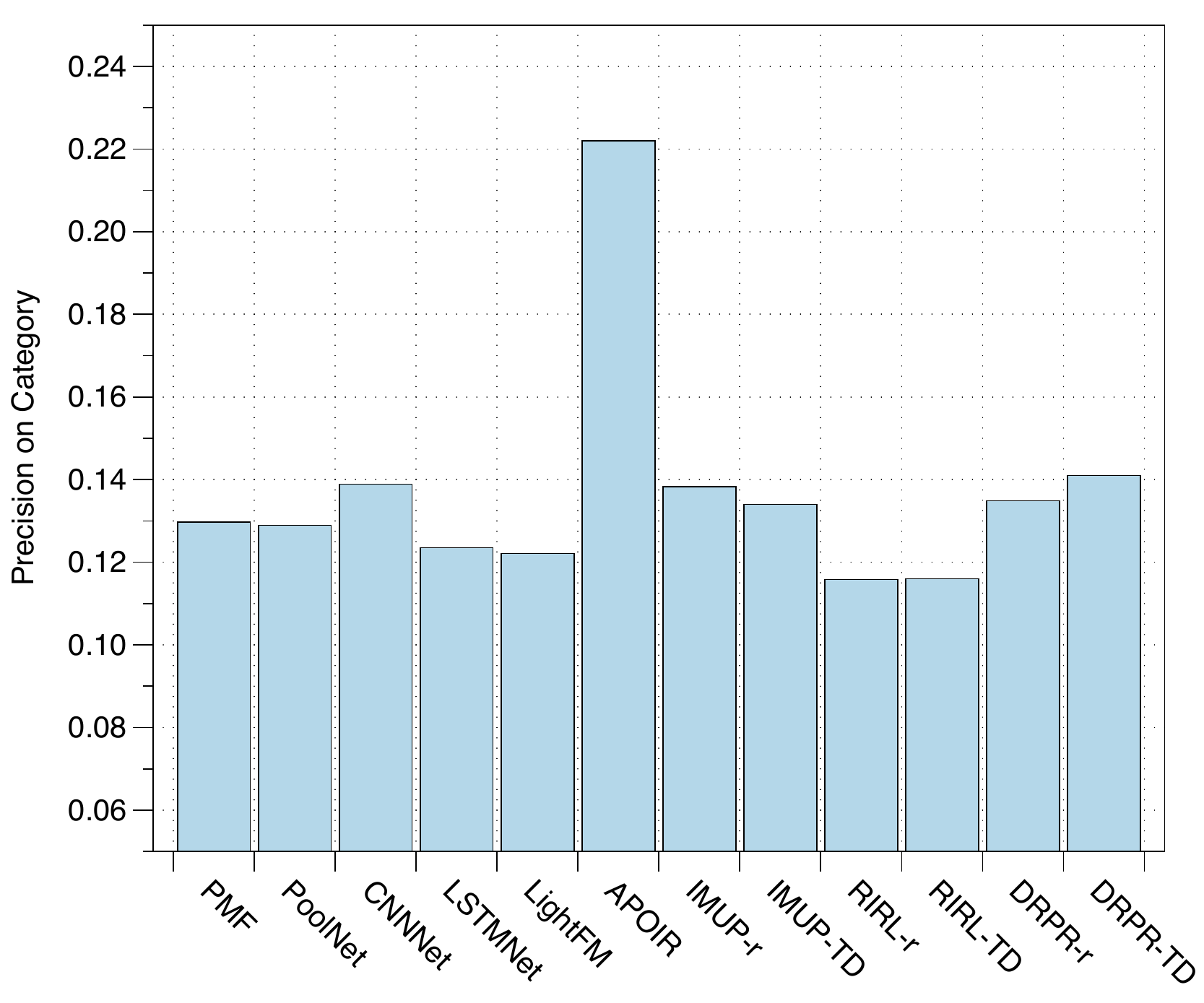}}
	\subfigure[Recall on Category]{\label{fig:overall_rec_cat_tky}\includegraphics[width=4.35cm]{{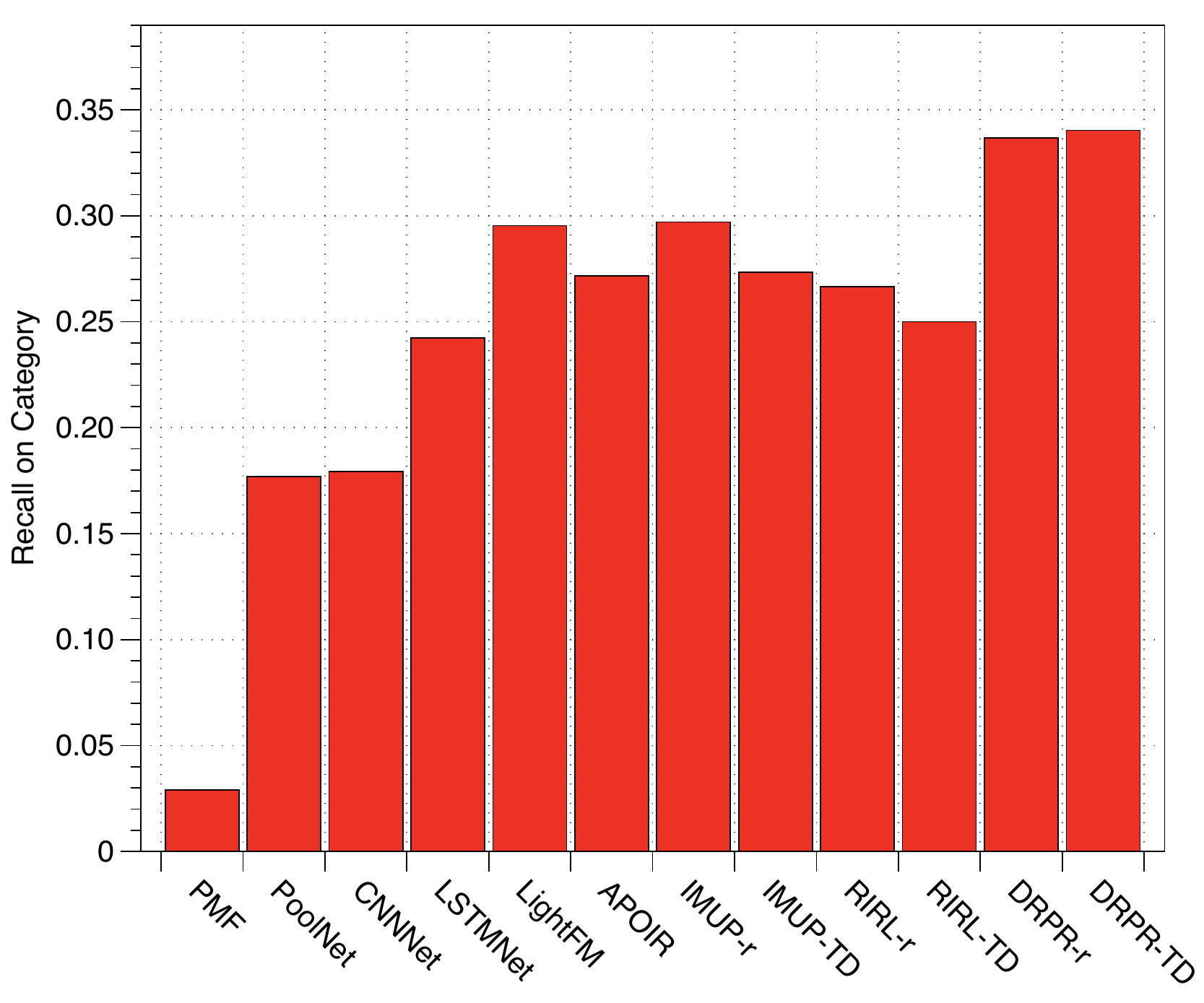}}}
	\subfigure[Average Similarity]{\label{fig:overall_avg_sim_tky}\includegraphics[width=4.35cm]{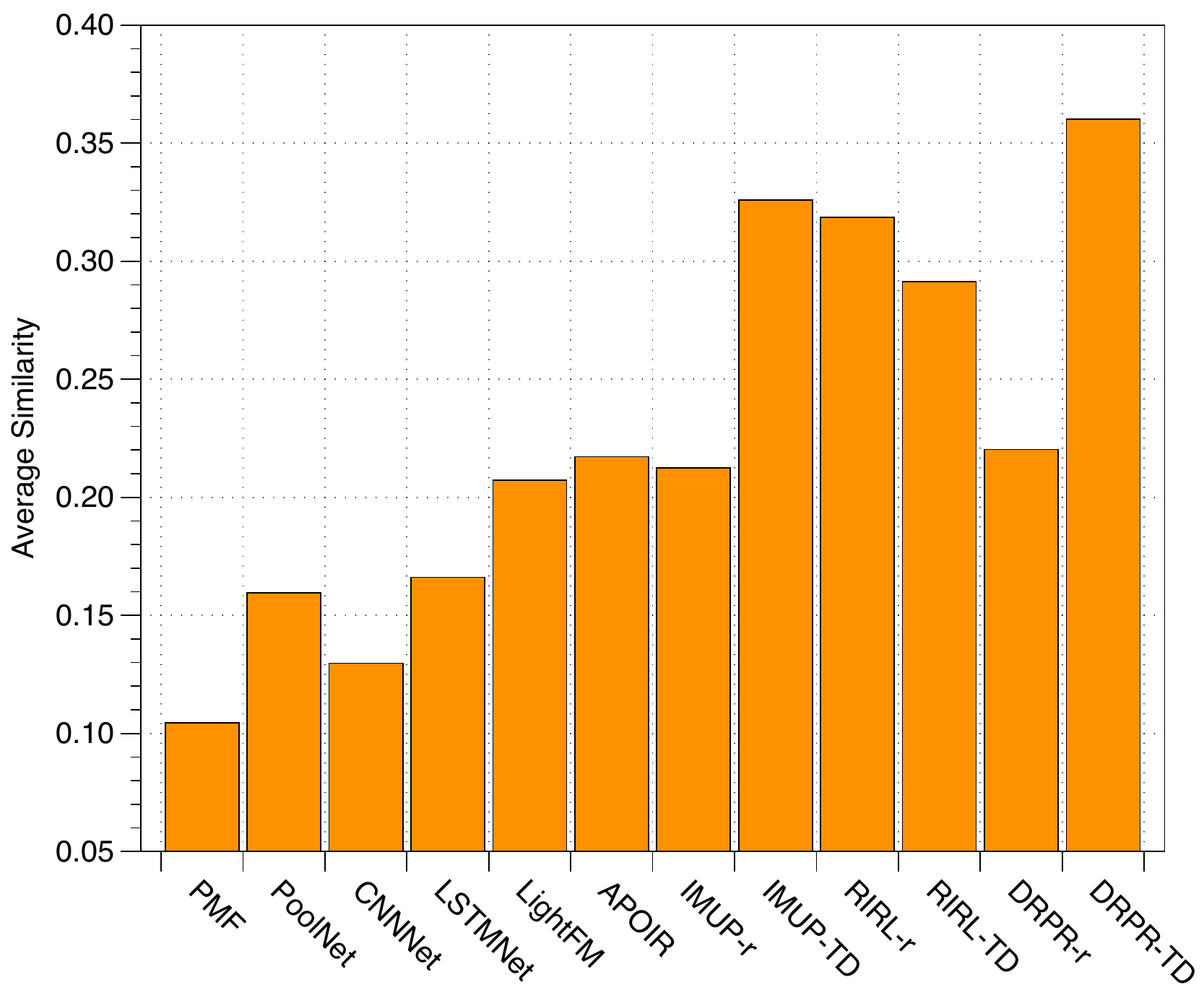}}
	\subfigure[Average Distance]{\label{fig:overall_avg_dis_tky}\includegraphics[width=4.35cm]{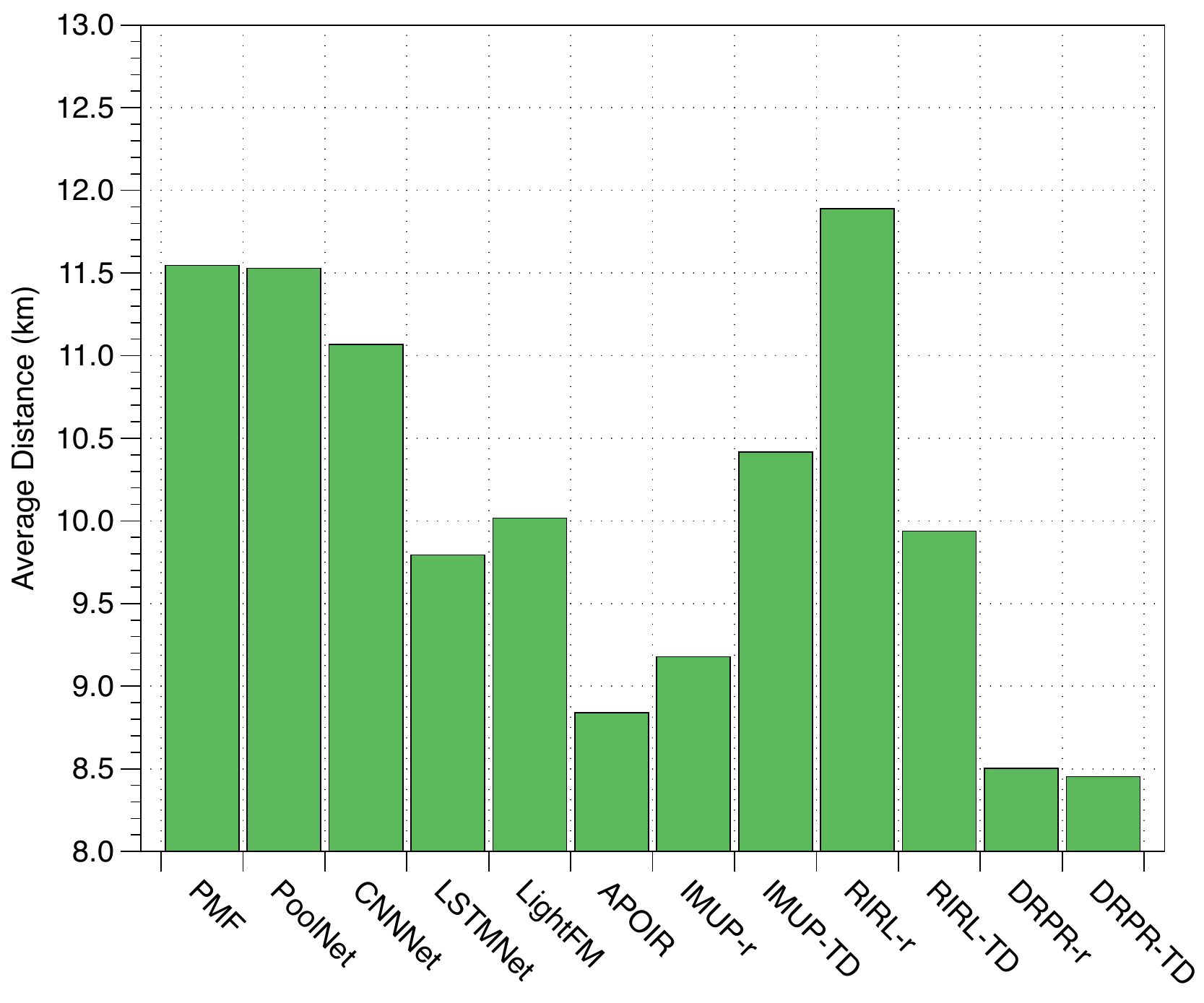}}
		\vspace{-0.3cm}
	\captionsetup{justification=centering}
	\caption{Overall performance {\it w.r.t.} Tokyo dataset.}
		\vspace{-0.1cm}
	\label{fig:tky_overall}
\end{figure*}

\begin{table*}[!ht]
\renewcommand\arraystretch{1}
	\centering
	
	\caption{Ablation Study of DRPR {\it w.r.t.} New York dataset }
		\vspace{-0.2cm}
	\begin{tabular}{c|cc|cc|cc|cc|c}
		\hline
		           & Prec\_Cat     & Outperform & Rec\_Cat     & Outperform & Avg\_Sim     & Outperform & Avg\_Dist     & Outperform & Time Cost\\ \hline

DRPR        & 0.1086 & $-$  & 0.0913 & $-$ & 0.3890 & $-$  & 11.351 & $-$ & 32.760 \\
DRPR$^*$    & 0.0313 & $+71.2\%$  & 0.0184 & $+79.8\%$  & 0.3352 & $+13.8\%$ & 10.171 & $+10.4\%$ & -\\
DRPR$^{\prime}$   & 0.0359 & $+66.9\%$  & 0.0270 & $+70.4\%$  & 0.3362 & $+13.6\%$  & 11.855 & $+4.44\%$ & - \\
DRPR$^-$        & 0.0201 & $+81.5\%$     & 0.0100 & $+89.0\%$  & 0.3283 & $+15.6\%$  & 12.591 & $+10.9\%$ & 53.236   \\
\hline
	\end{tabular}
	\vspace{-0.1cm}
	\label{tab:ablation_study_nyc}
\end{table*}

\begin{table*}[!ht]
\renewcommand\arraystretch{1}
	\centering
	\caption{Ablation Study of DRPR {\it w.r.t.} Tokyo dataset }
		\vspace{-0.2cm}
	\begin{tabular}{c|cc|cc|cc|cc|c}
		\hline
		           & Prec\_Cat     & Outperform & Rec\_Cat     & Outperform & Avg\_Sim     & Outperform & Avg\_Dist     & Outperform & Time Cost\\ \hline

DRPR        & 0.1409 & $-$  & 0.3403 & $-$ & 0.2202 & $-$  & 8.4533 & $-$ & 31.837  \\
DRPR$^*$    & 0.1266 & $+10.1\%$  & 0.3234 & $+4.97\%$  & 0.2193 & $+0.41\%$ & 8.7521 & $+3.53\%$ & - \\
DRPR$^{\prime}$   & 0.1289 & $+8.52\%$  & 0.3243 & $+4.70\%$  & 0.2171 & $+1.41\%$  & 10.055 & $+18.9\%$ & - \\
DRPR$^-$        & 0.1204 & $+14.5\%$     & 0.0353 & $+89.6\%$  & 0.1899 & $+13.8\%$  & 9.5301 & $+12.7\%$ & 58.591  \\
\hline
	\end{tabular}
	\label{tab:ablation_study_tky}
	\vspace{-0.2cm}
\end{table*}

\subsubsection{Evaluation Metrics}
We evaluated  performance in terms of four metrics:

\noindent {\bf (1) $\text{Precision on Category (Prec\_Cat)}$.}
POI recommendation on the POI category level can be viewed as multi-classification. We used the weighted precision, denoted by:
\begin{equation}
    \text{Prec\_Cat} = \frac{|c_k| \cdot I_{TP}^k}{\sum \limits_k |c_k| (I_{TP}^k+I_{FP}^k)}
\end{equation}
where $c_k$ is the $k$-th POI category,  $|c_k|$ is the number of $c_k$, $I_{TP}^k$ is the number of true positive predictions, and $I_{FP}^k$ is the number of false positive predictions.

\noindent {\bf (2) $\text{Recall on Category (Rec\_Cat)}$.}
We used the weighted recall on POI categories, denoted by:
\begin{equation}
    \text{Rec\_Cat} = \frac{|c_k| \cdot I_{TP}^k}{\sum \limits_k |c_k| (I_{TP}^k+I_{FN}^k)}
\end{equation}
where  $I_{FN}^{k}$ is  number of false negative predictions for $c_k$.

\noindent {\bf (3) $\text{Average Similarity (Avg\_Sim)}$.}
We expected that the semantic meaning of the predicted POI category should be similar to the actual goal of the user traveling.
Thus, we evaluated the average similarity between the real and predicted POI category.
We employed the pretrained {\it Glove} word embedding~\cite{pennington2014glove} to represent POI categories,
then calculated the cosine similarity between the real POI category ``word$^l$'' and the predicted POI category ``$\hat{\text{word}}^{l}$''.
Formally, the average similarity is given by: 
\begin{equation}
    \text{Avg\_Sim} = \frac{\sum \limits_{l} cosine(\text{word}^l, \hat{\text{word}}^{l})}{L}
\end{equation}
where $L$ denotes the total visit number. 
The higher value of Avg\_Sim is, the better the model performance is.

\noindent {\bf (4) $\text{Average Distance (Avg\_Dist)}$.}
We evaluated the average geographic distance between the locations of predicted POIs and real visit POIs.
Avg\_Dist can be defined as follows:
\begin{equation}
    \text{Avg\_Dist} = \frac{\sum \limits_{l} Dist(P^{l}, \hat{P}^{l})}{L}.
\end{equation}
where $Dist(P^{l}, \hat{P}^{l})$ denotes the distance between the location of the real POI $P^l$ and the predicted POI $\hat{P}^l$ at the $l$-th visit.
The lower value of Avg\_Dist is, the better the model performance is.

\begin{figure*}[!thb]
	\centering
	\subfigure[Precision on Category]{\label{fig:robust_prec_cat_nyc}\includegraphics[width=4.35cm]{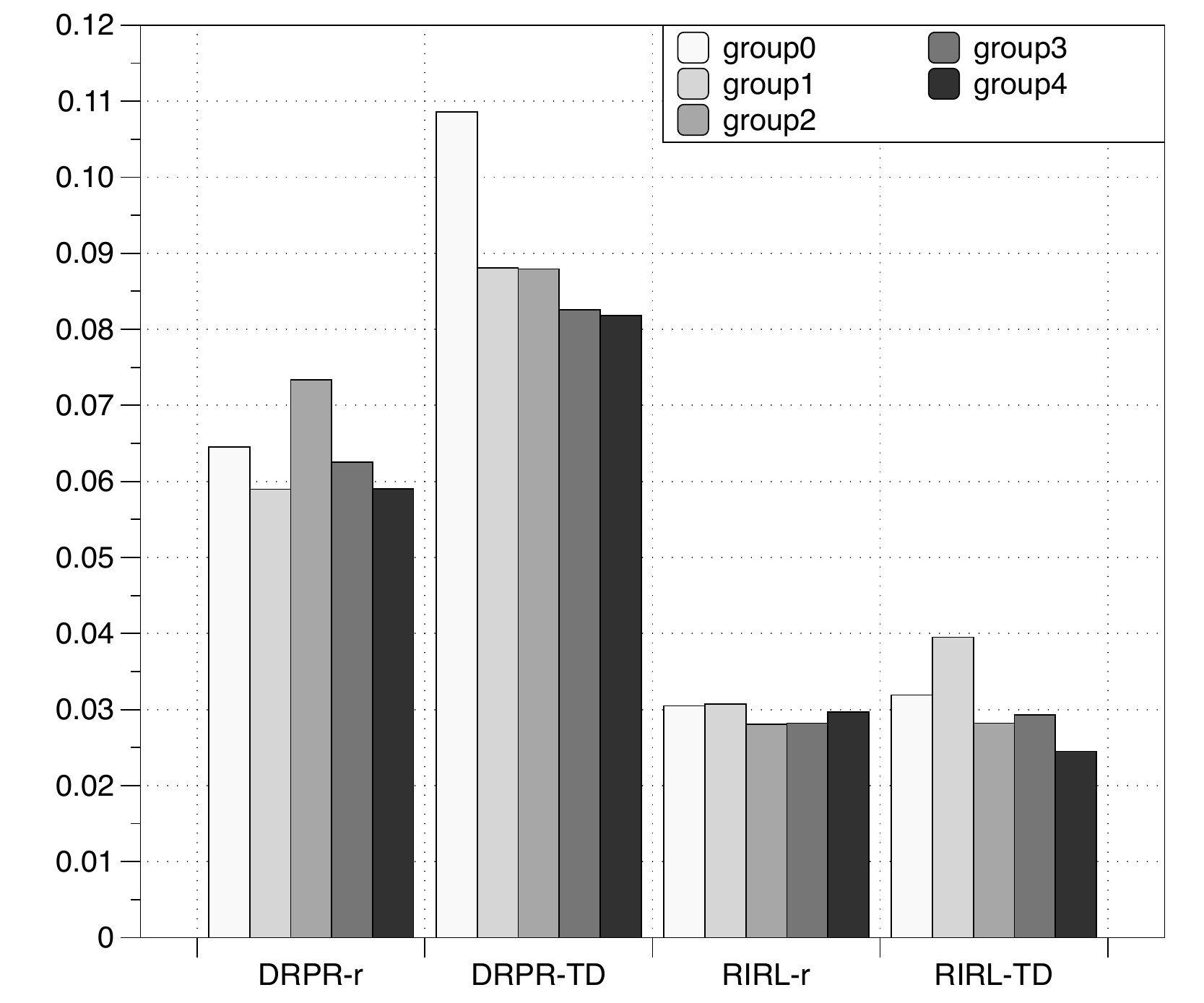}}
	\subfigure[Recall on Category]{\label{fig:robust_rec_cat_nyc}\includegraphics[width=4.35cm]{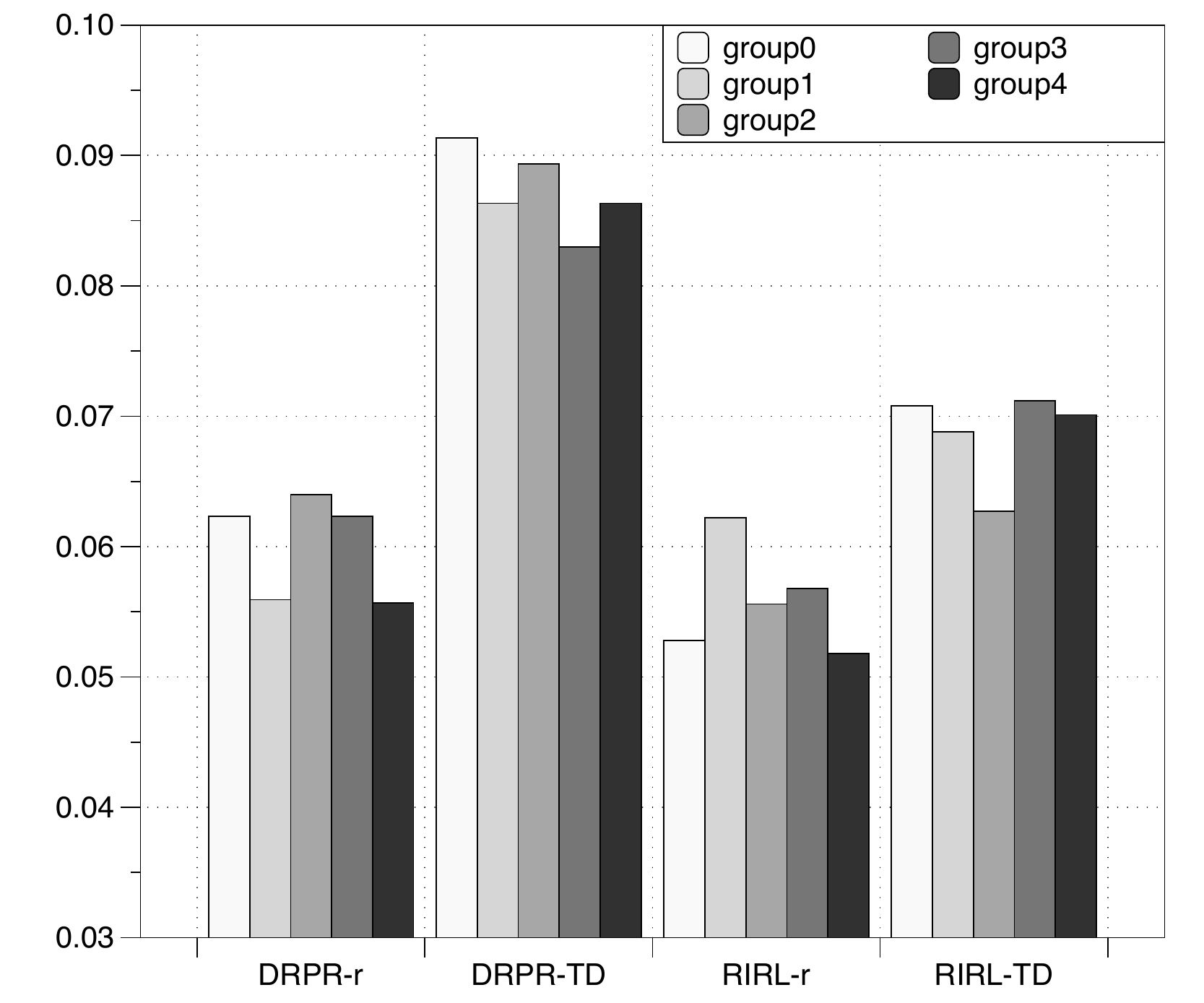}}
	\subfigure[Average Similarity]{\includegraphics[width=4.35cm]{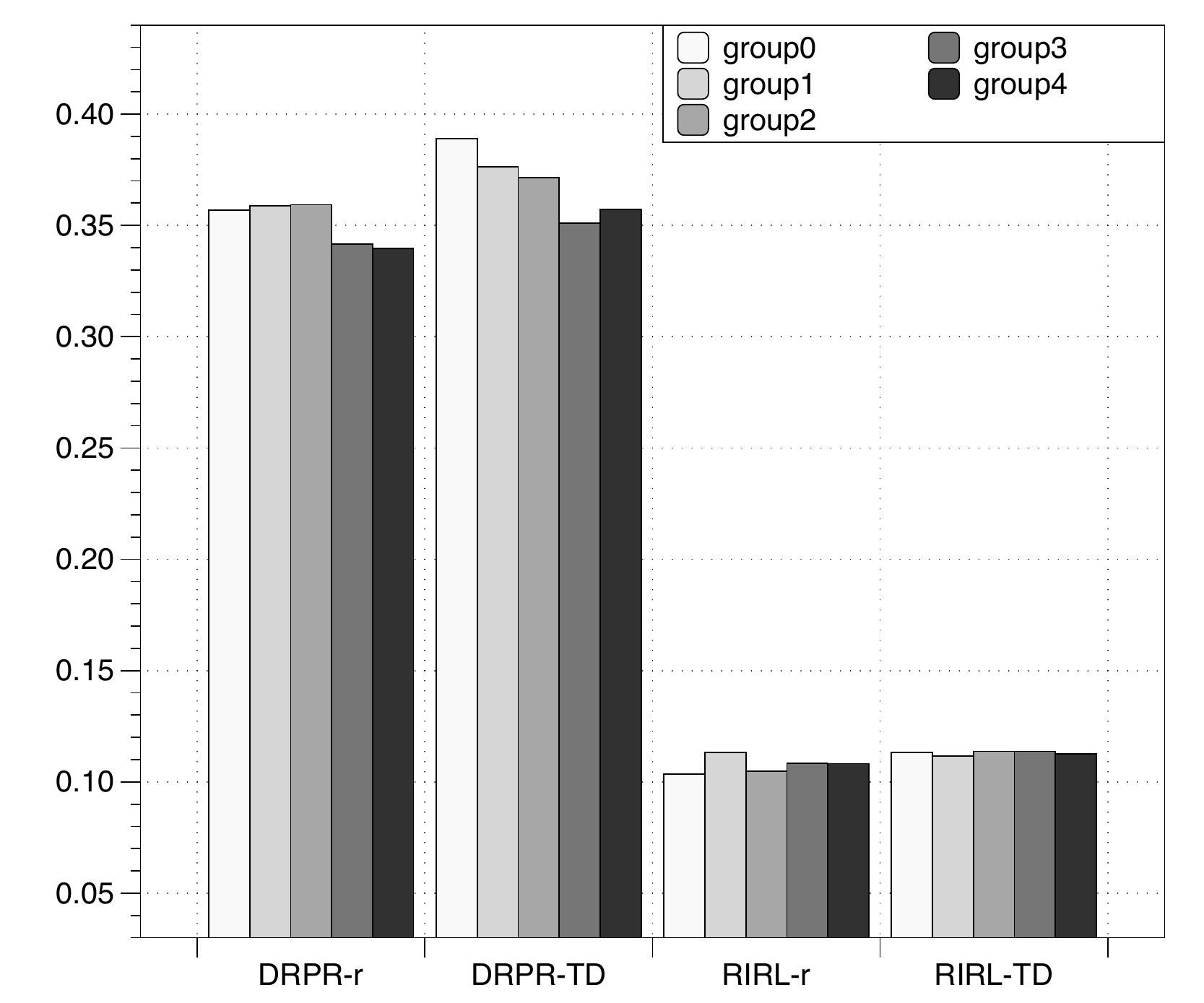}}
	\subfigure[Average Distance]{\label{fig:robust_avg_dis_nyc}\includegraphics[width=4.35cm]{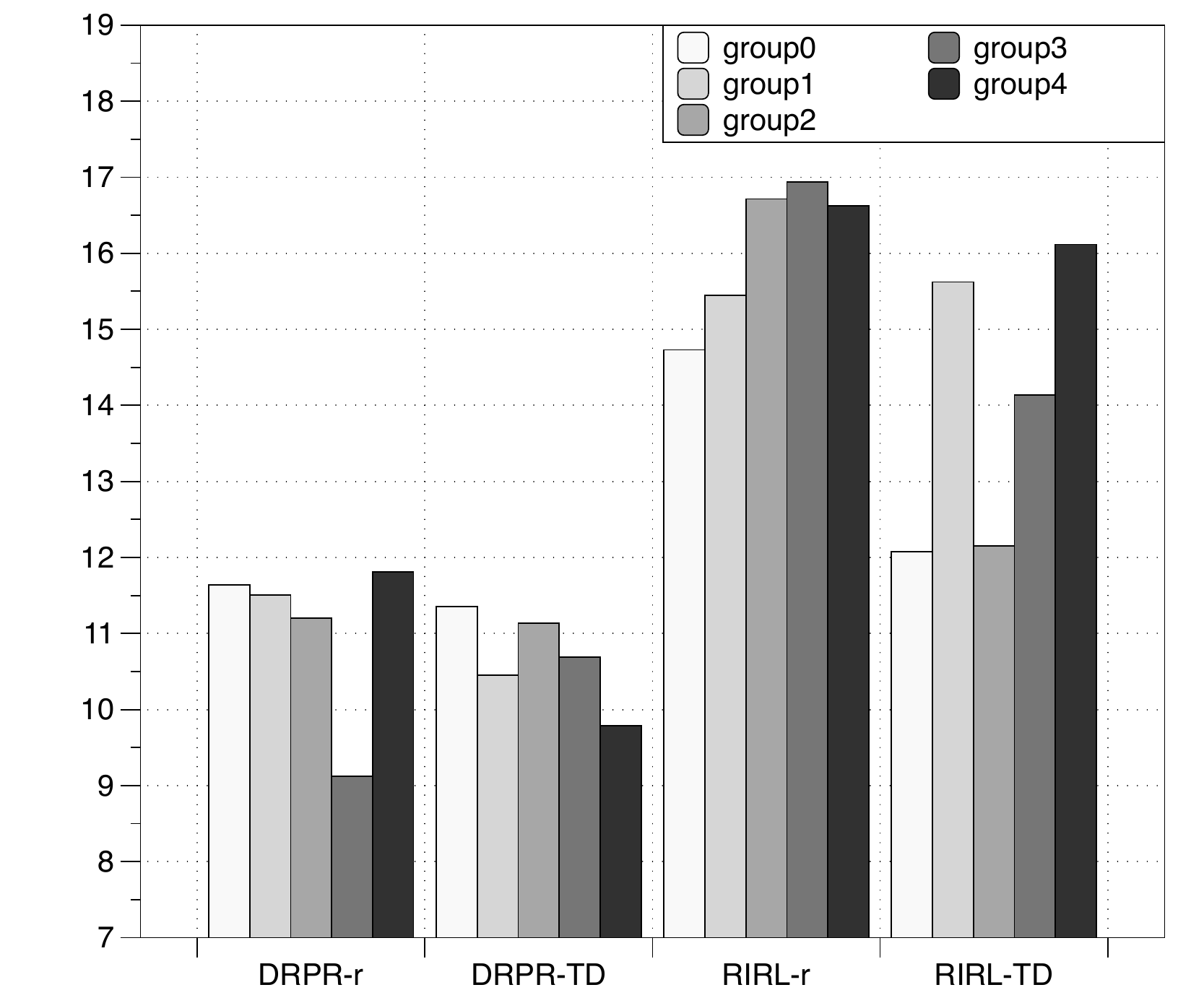}}
		\vspace{-0.3cm}
	\captionsetup{justification=centering}
	\caption{Robustness Check {\it w.r.t.} New York dataset.}
		\vspace{-0.5cm}
	\label{fig:nyc_robust}
\end{figure*}

\begin{figure*}[!thb]
	\centering
	\subfigure[Precision on Category]{\label{fig:robust_prec_cat_tky}\includegraphics[width=4.35cm]{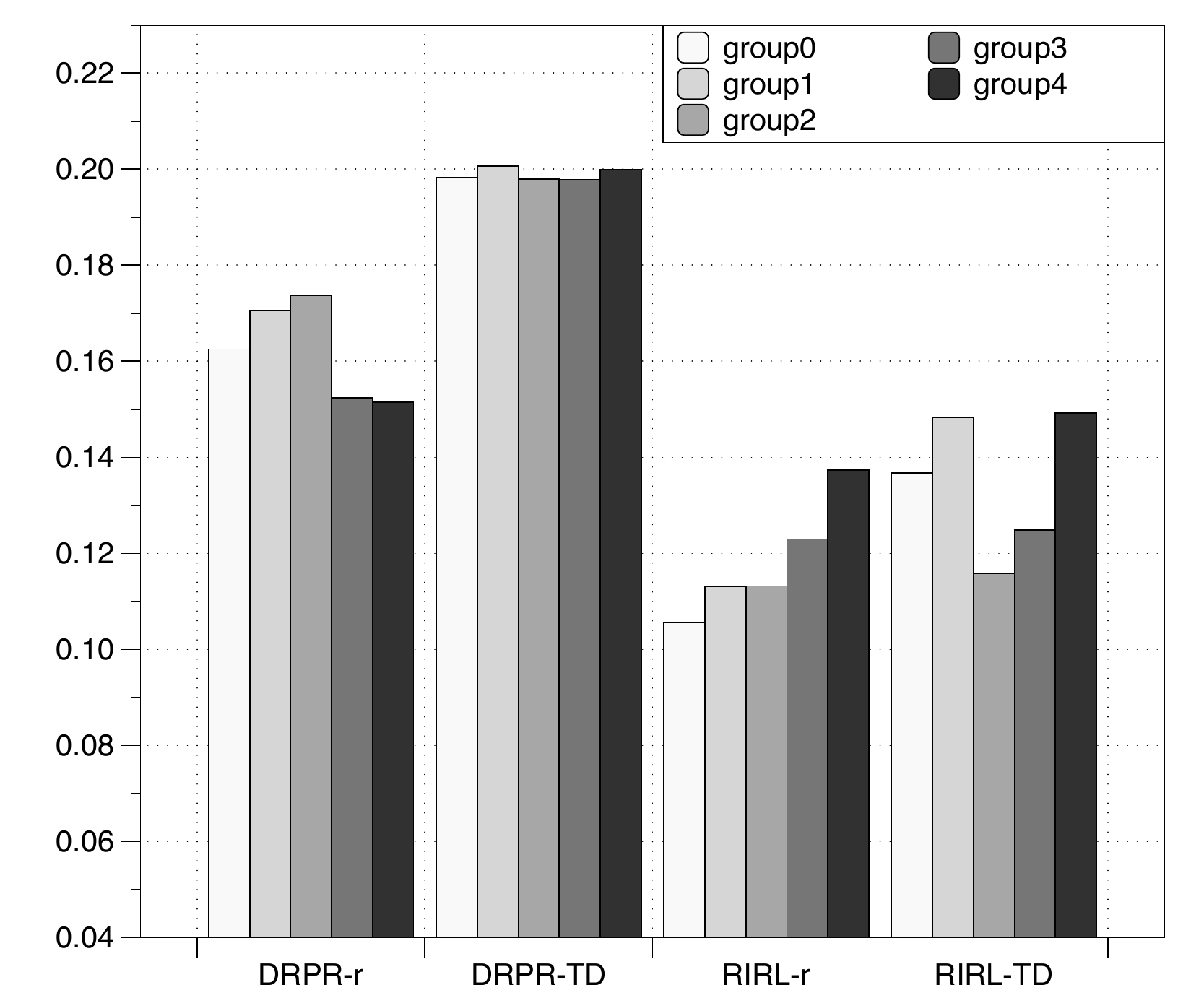}}
	\subfigure[Recall on Category]{\label{fig:robust_rec_cat_tky}\includegraphics[width=4.35cm]{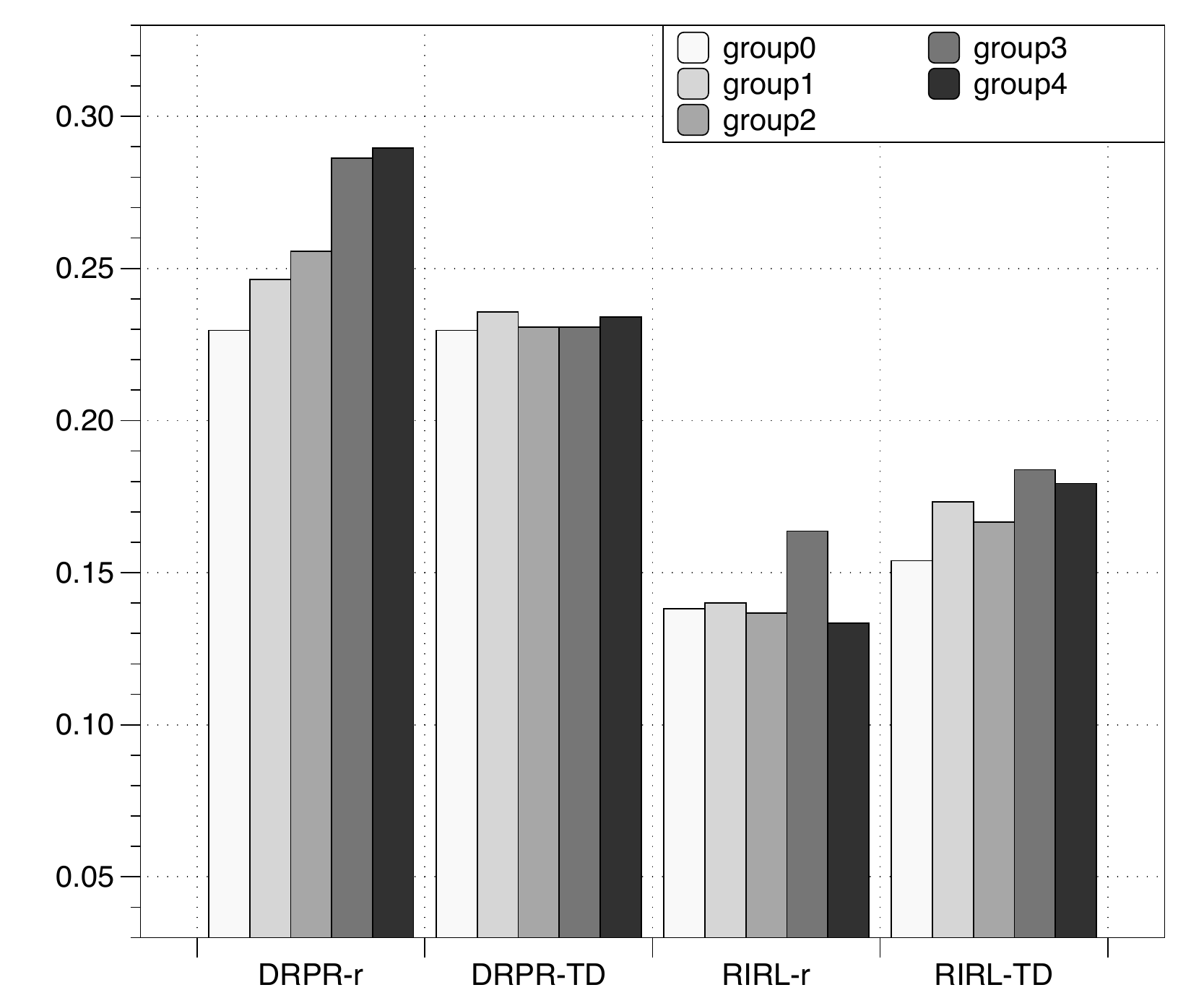}}
	\subfigure[Average Similarity]{\label{fig:robust_avg_sim_tky}\includegraphics[width=4.35cm]{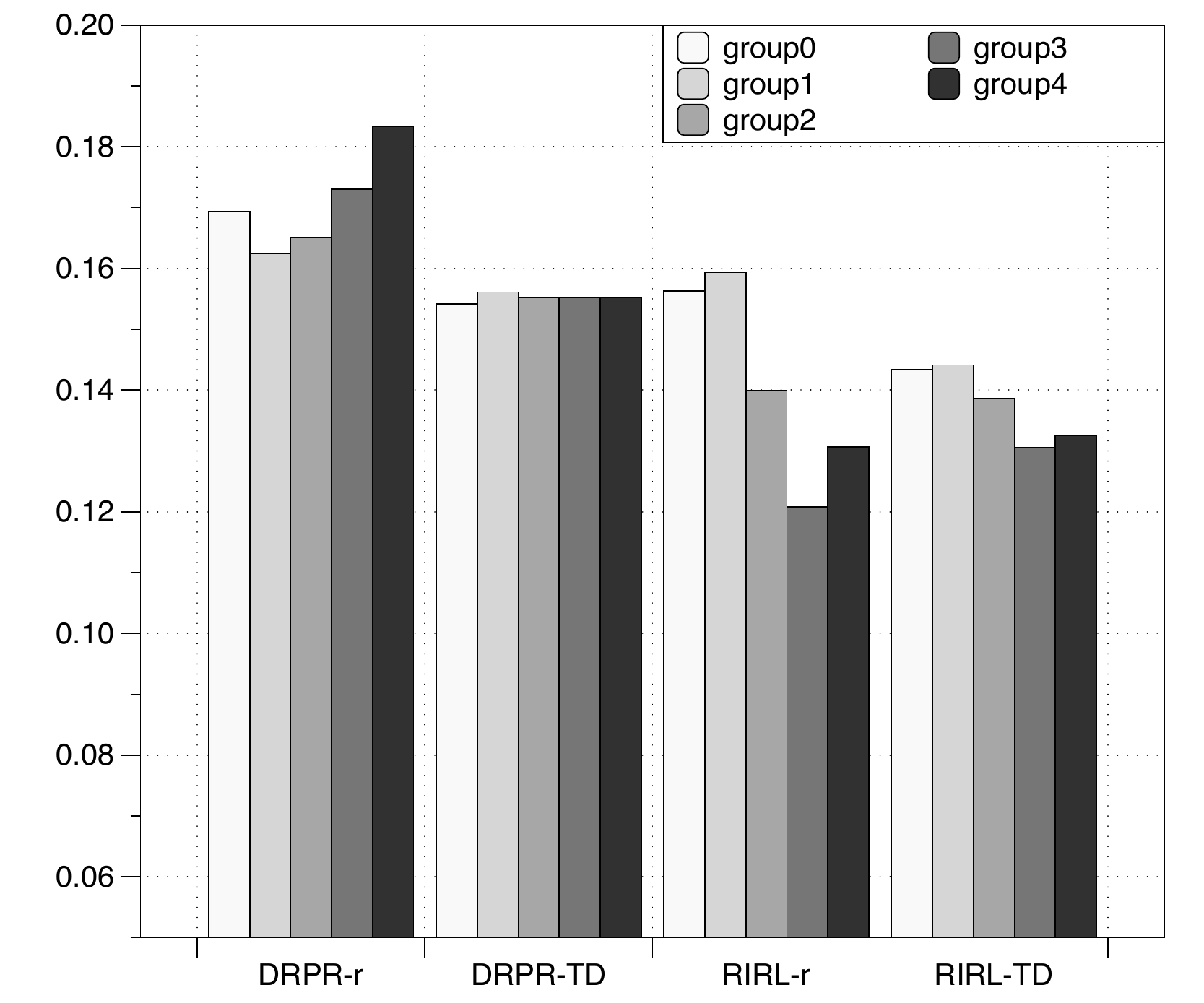}}
	\subfigure[Average Distance]{\label{fig:robust_avg_dis_tky}\includegraphics[width=4.35cm]{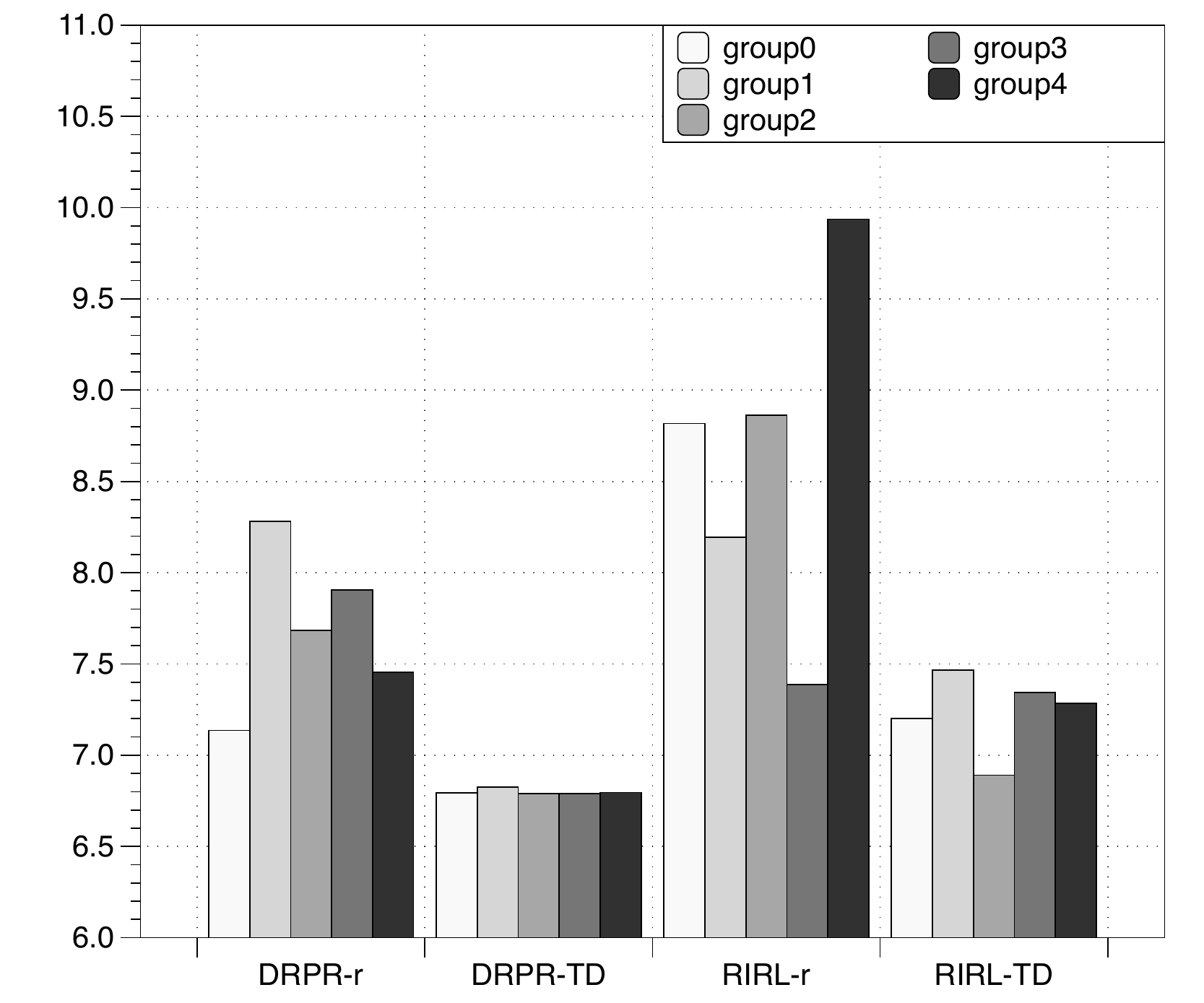}}
		\vspace{-0.3cm}
	\captionsetup{justification=centering}
	\caption{Robustness Check {\it w.r.t.} Tokyo dataset.}
		\vspace{-0.4cm}
	\label{fig:tky_robust}
\end{figure*}

\subsubsection{Baseline Models}
We compared the performance of our enhanced POI recommendation framework (namely ``DRPR'',  \underline{D}ynamic \underline{R}einforced \underline{P}OI \underline{R}ecommendation) against the following baseline algorithms.
{\bf (1) PMF} recommends items based on the user-item interaction matrix through probabilistic matrix factorization ~\cite{mnih2008probabilistic}. 
{\bf (2) PoolNet} employs a deep neural network model to model the interaction between user embedding and item embedding for recommending items ~\cite{covington2016deep}.
{\bf (3) WaveNet} is used to generate raw audio waveforms originally. 
It also can be used to recommend items by imitating the sequential decision-making process of users ~\cite{oord2016wavenet}.
{\bf (4) LSTMNet} utilizes a recurrent neural network to mimic users' behavior for recommending items ~\cite{hidasi2015session}.
{\bf (5) LightFM} utilizes user-item interaction matrix and user-item meta-data for recommending items ~\cite{DBLP:conf/recsys/Kula15}.
{\bf (6) APOIR} proposes an adversarial POI recommendation framework, which learns the distribution of user latent  preference via a minimax game optimization ~\cite{zhou2019adversarial}.
{\bf (7) IMUP-r} is a new POI recommendation framework with incorporating spatial KG and reinforcement learning to recommend items for users.
The reward-based sampling strategy is used to improve model performance ~\cite{wang2020incremental}.
{\bf (8) IMUP-TD} is the same as the model structure of the IMUP-r. 
The only difference is that the sampling strategy of IMUP-TD is TD-based.
{\bf (9) RIRL-r} is our preliminary work, which utilizes the adversarial training skill to train the whole framework and employs the reward-based sampling strategy during the training phase ~\cite{DBLP:conf/aaai/WangW0ZHF21}.
{\bf (10) RIRL-TD} is a variant of RIRL-r, which utilizes TD-based sampling strategy during the training phase.

\subsubsection{Hyperparameters, Source Code, and Reproducibility.}
In the experiment, we first selected the 15,000 continuous check-in records from New York and Tokyo datasets respectively.
Then, for each city, we split the corresponding records into two  non-overlapping sets:
the earliest $80\%$  for training and the remaining $20\%$ for testing.
During the learning process,  we set the dimension of the state as 200 and the number of POI candidates as 20.
For the implementation of baseline models,
we evaluated PMF by adopting the implementation  
\footnote{\url{https://github.com/fuhailin/Probabilistic-Matrix-Factorization}}; 
we implemented PoolNet, WaveNet, and LSTMNet by adopting ``spotlight'' ~\cite{kula2017spotlight}, where the learning rate is set as 0.1;
we evaluated LightFM by adopting the implementation \footnote{\url{https://github.com/lyst/lightfm}};
we implemented IMUP by adopting the implementation 
\footnote{\url{https://github.com/wangdongjie100/KDD2020}}.
Finally, to make others reproduce the experiment easily, we release the code and data in Dropbox\footnote{\url{https://www.dropbox.com/sh/7irkv3ar8xd1rr0/AADfUQR5vz6hbuzoayu-\_DQQa?dl=0}}.

\subsubsection{Environmental Settings.}
We conducted all experiments on Ubuntu 18.04.3 LTS, Intel(R) Core(TM) i9-9920X CPU@ 3.50GHz, with Titan RTX and memory size 128G.
In addition, we implemented all models based on python 3.7.4, TensorFlow 2.0.0 GPU.

\vspace{-0.2cm}
\subsection{Overall Performance (\textbf{Q1})}
We compared our method (DRPR) with baseline algorithms in terms of Prec\_Cat, Rec\_Cat, Avg\_Sim and Avg\_Dist.
Figure \ref{fig:nyc_overall} and Figure \ref{fig:tky_overall} show that our method outperforms other baseline models on both New York and Tokyo datasets.
A potential interpretation for the improvement on Prec\_Cat and Rec\_Cat is that DRPR sufficiently captures the visit preferences of users by mining the sub-structure information in the dynamic KG. 
In addition, a possible reason for the improvement on Avg\_Sim and Avg\_Dist is that the evolving update strategy of the dynamic KG captures the change of user mobility patterns over time. 
Thus, not only from a semantic viewpoint, but also from a distance standpoint, DRPR achieves a high level of recommendation performance in comparison to baseline models.

\begin{figure*}[!thb]
	\centering
	\subfigure[Precision on Category]{\label{fig:reward_prec_cat_nyc}\includegraphics[width=4.35cm]{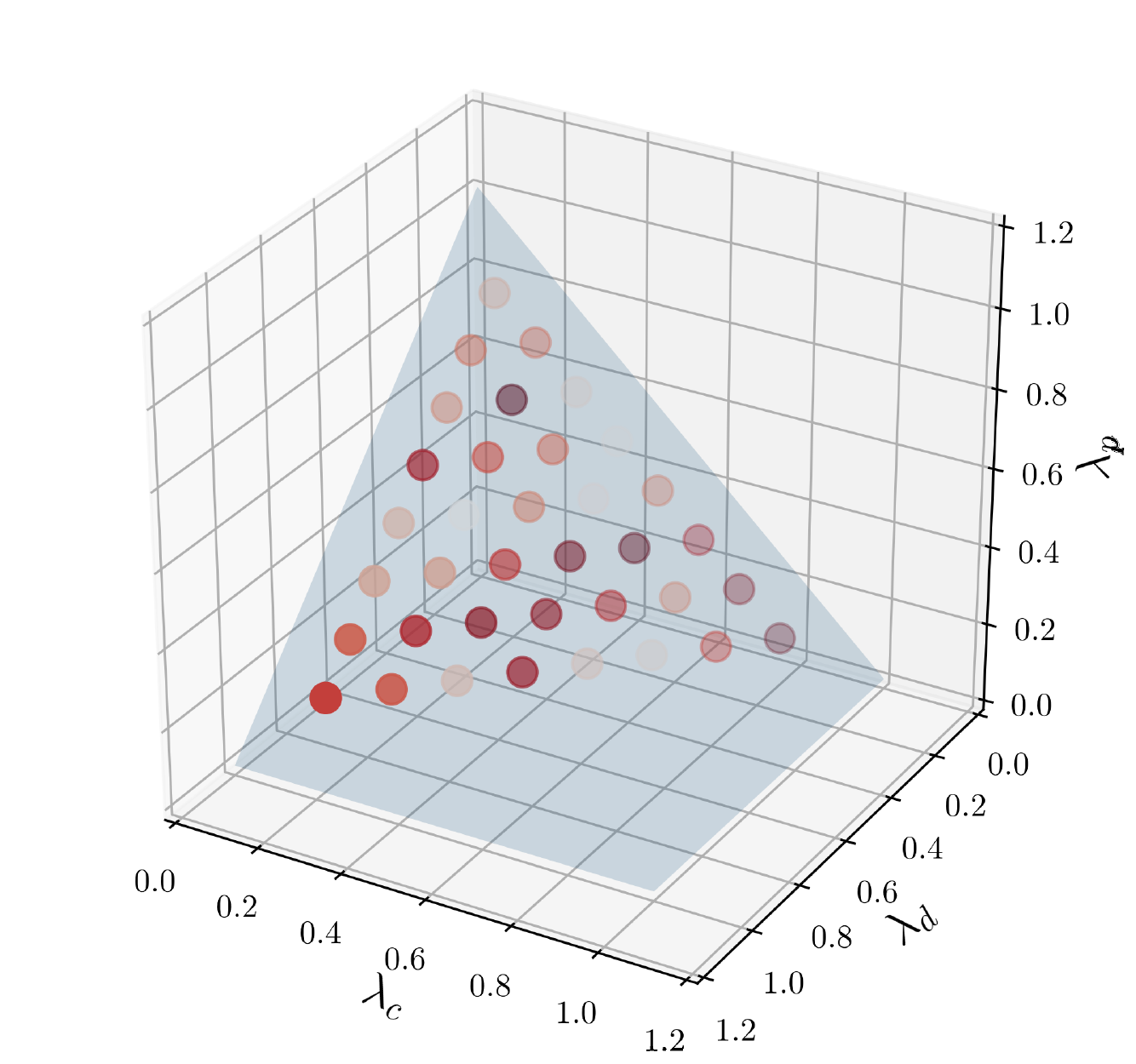}}
	\subfigure[Recall on Category]{\label{fig:reward_rec_cat_nyc}\includegraphics[width=4.35cm]{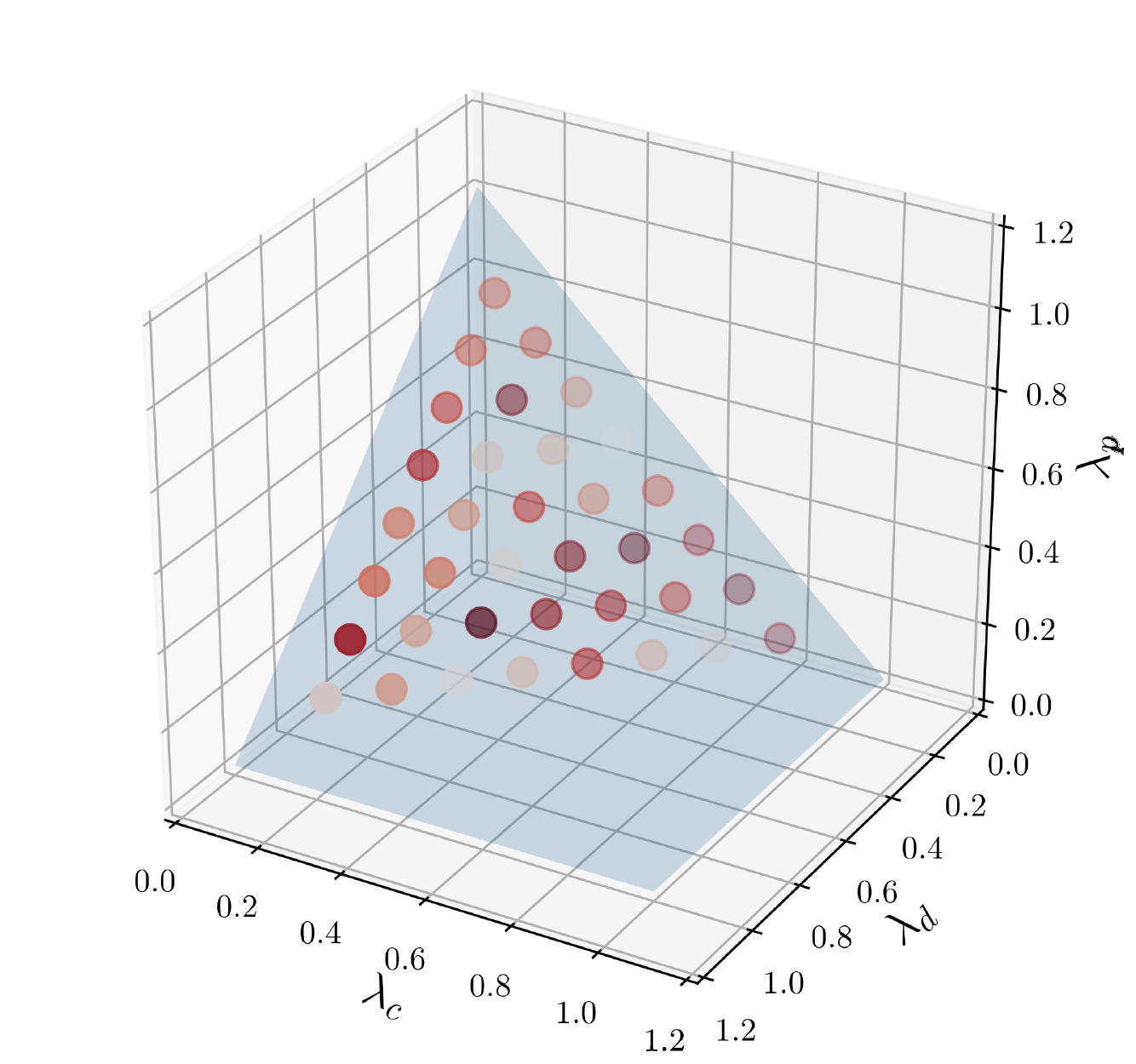}}
	\subfigure[Average Similarity]{\includegraphics[width=4.35cm]{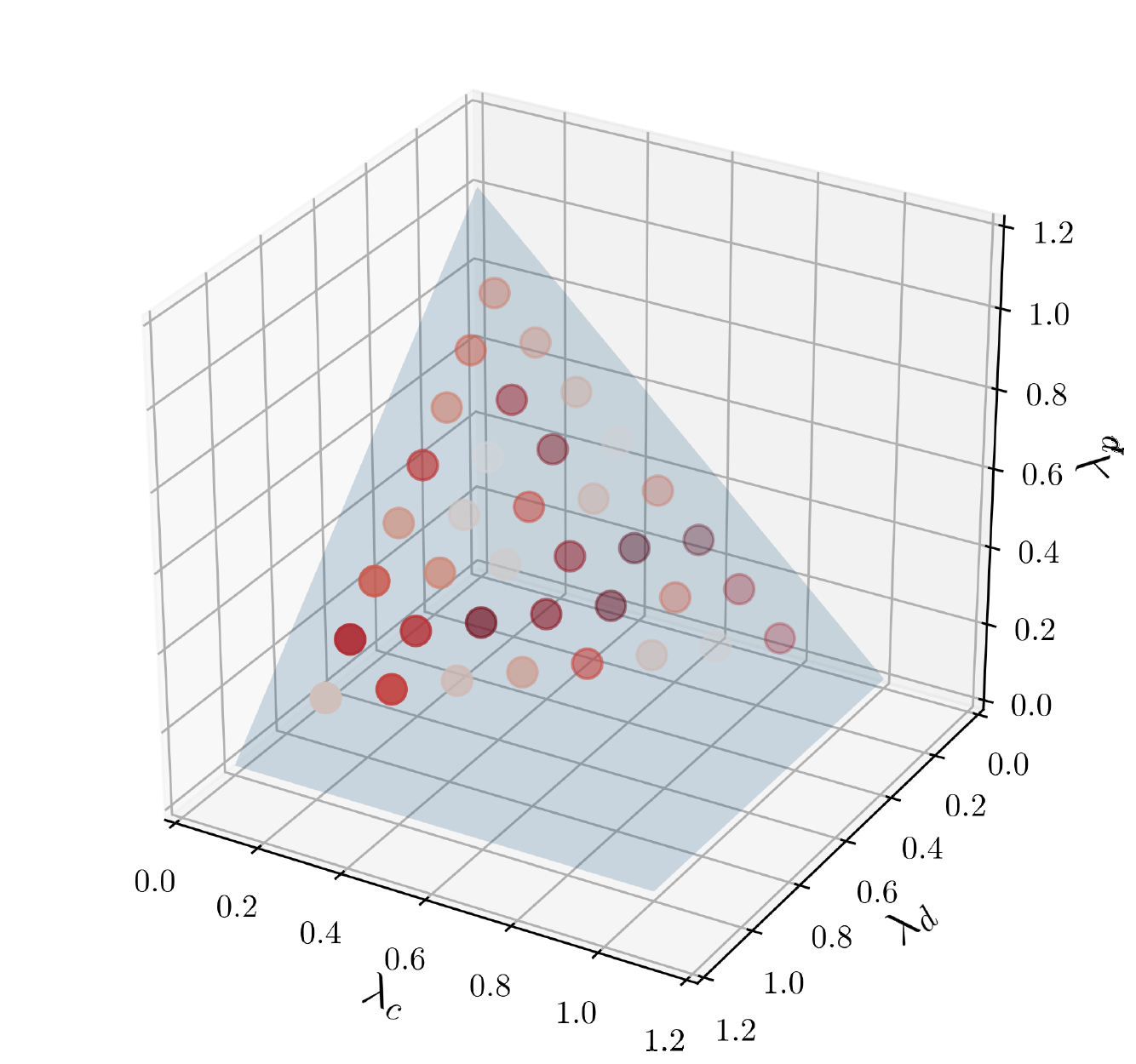}}
	\subfigure[Average Distance]{\label{fig:reward_avg_dis_nyc}\includegraphics[width=4.35cm]{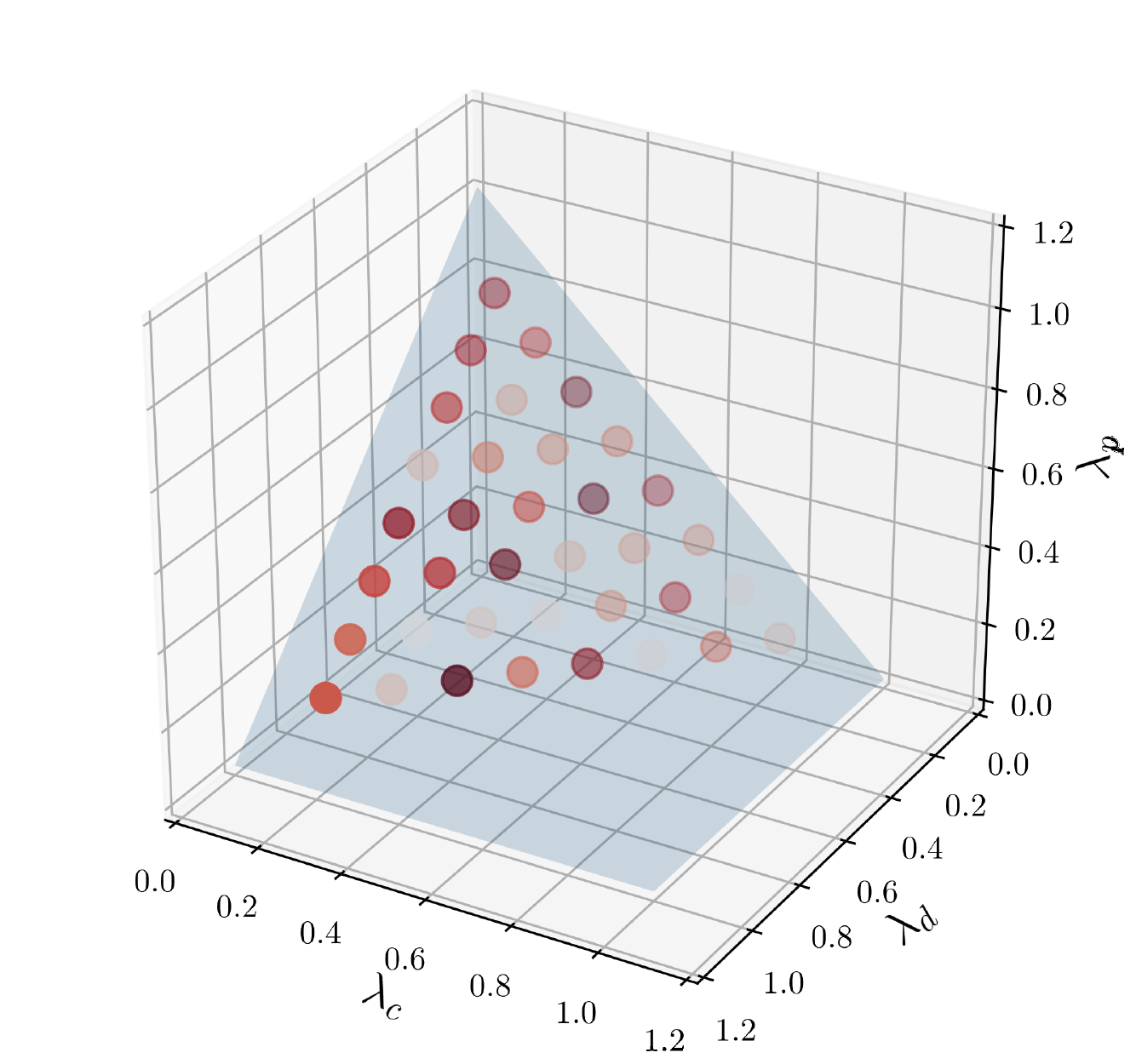}}
		\vspace{-0.3cm}
	\captionsetup{justification=centering}
	\caption{Reward Analysis {\it w.r.t.} New York dataset.}
		\vspace{-0.5cm}
	\label{fig:nyc_reward}
\end{figure*}

\begin{figure*}[!thb]
	\centering
	\subfigure[Precision on Category]{\label{fig:reward_prec_cat_tky}\includegraphics[width=4.35cm]{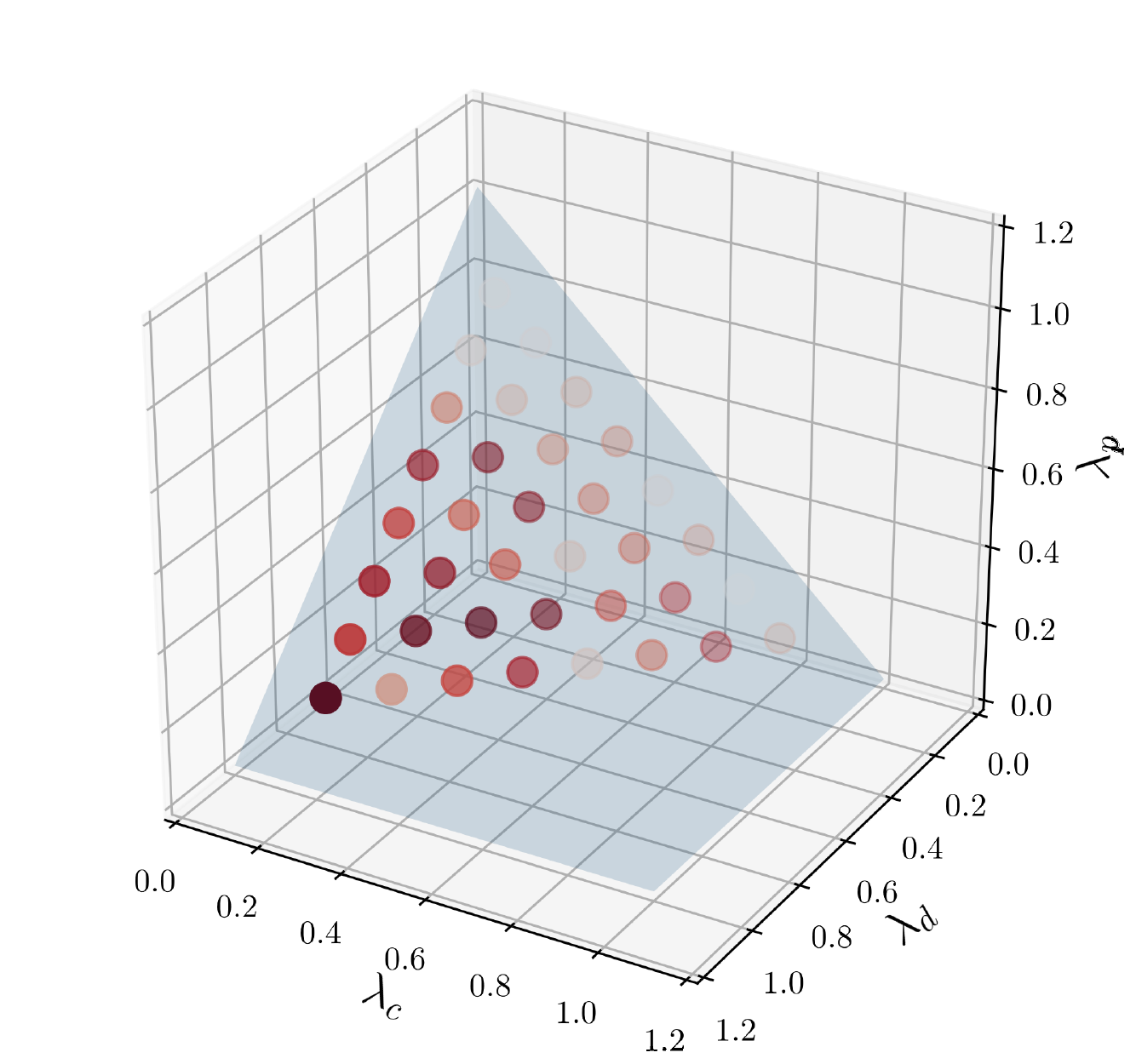}}
	\subfigure[Recall on Category]{\label{fig:reward_rec_cat_tky}\includegraphics[width=4.35cm]{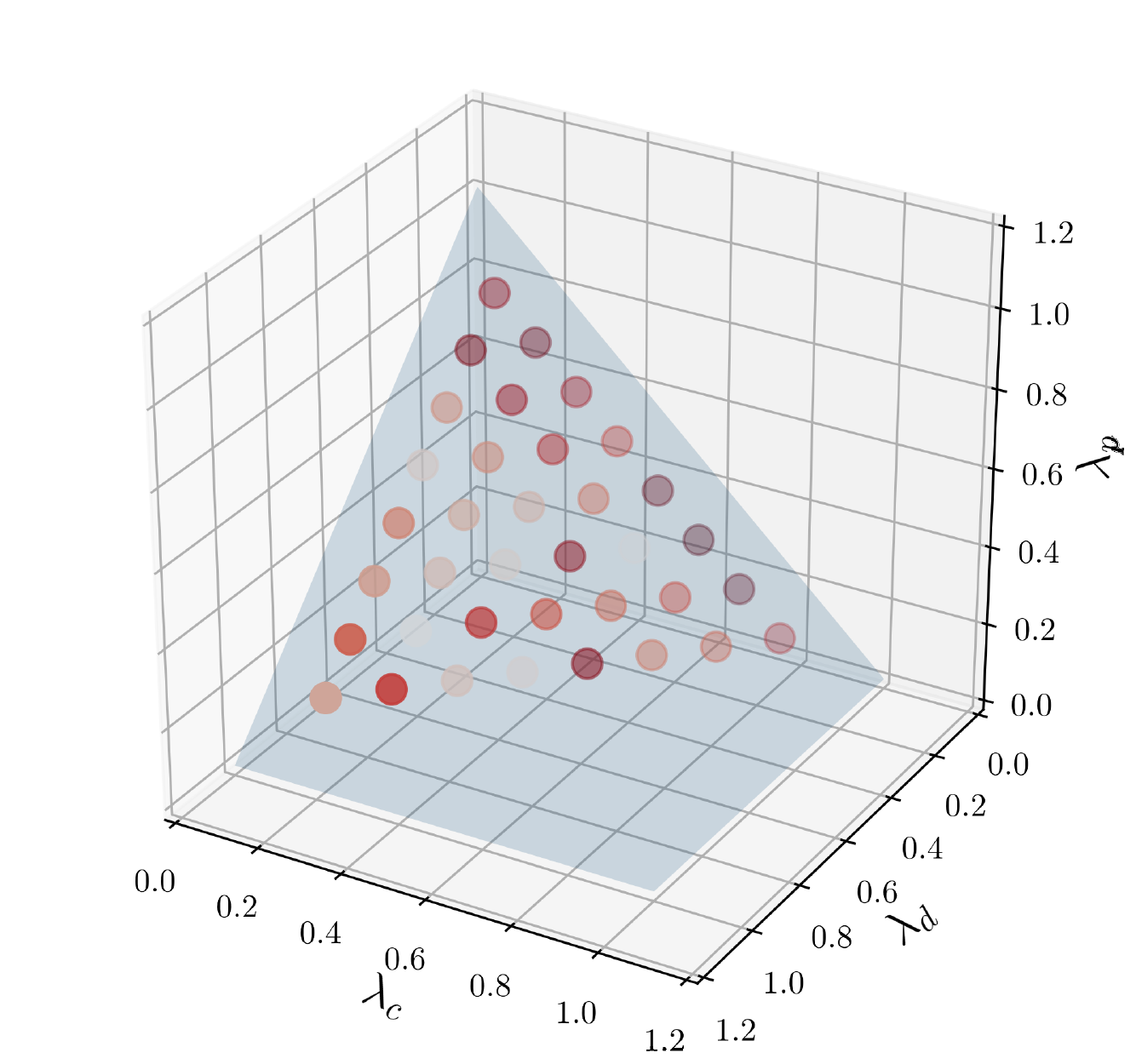}}
	\subfigure[Average Similarity]{\label{fig:reward_avg_sim_tky}\includegraphics[width=4.35cm]{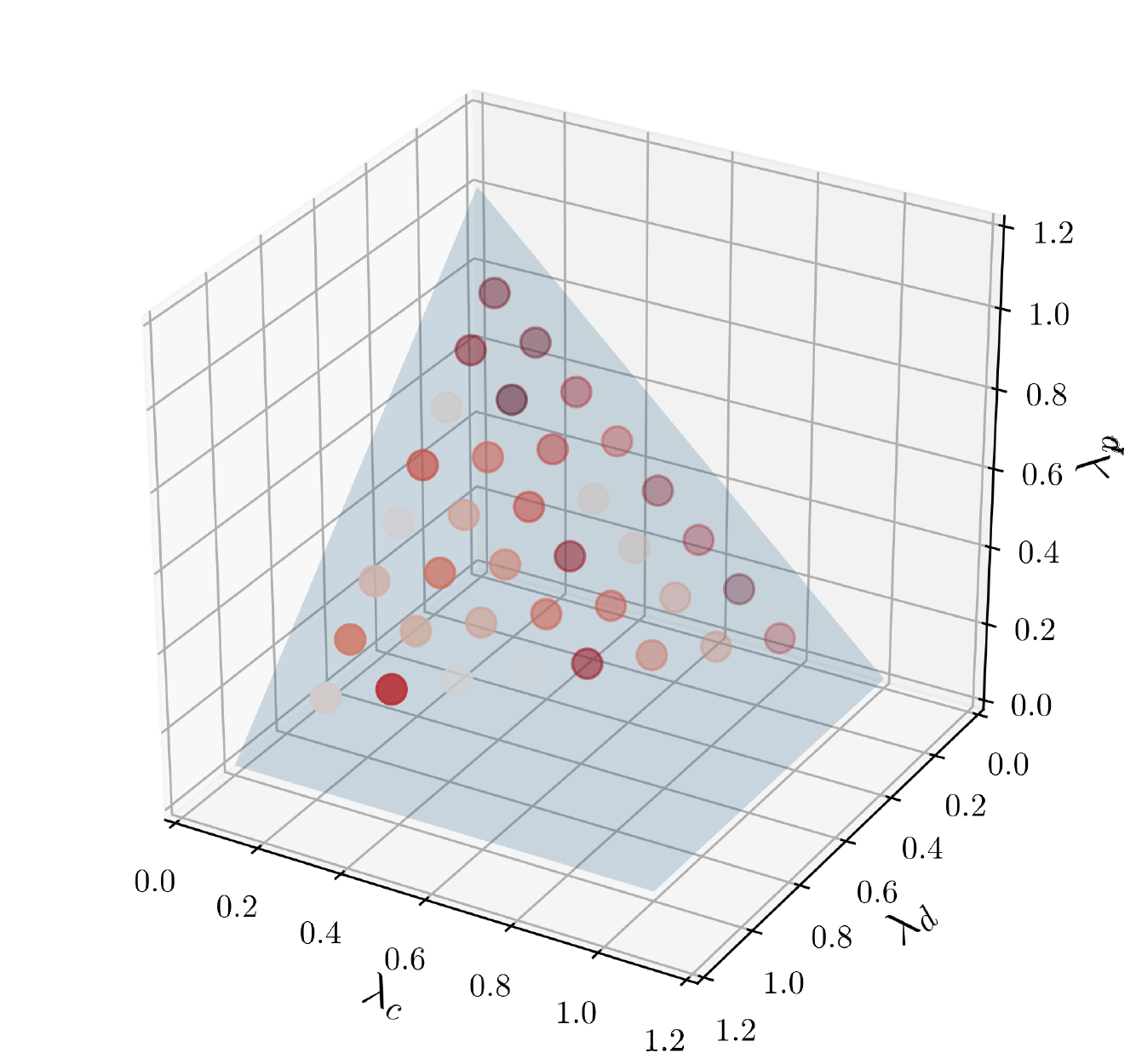}}
	\subfigure[Average Distance]{\label{fig:reward_avg_dis_tky}\includegraphics[width=4.35cm]{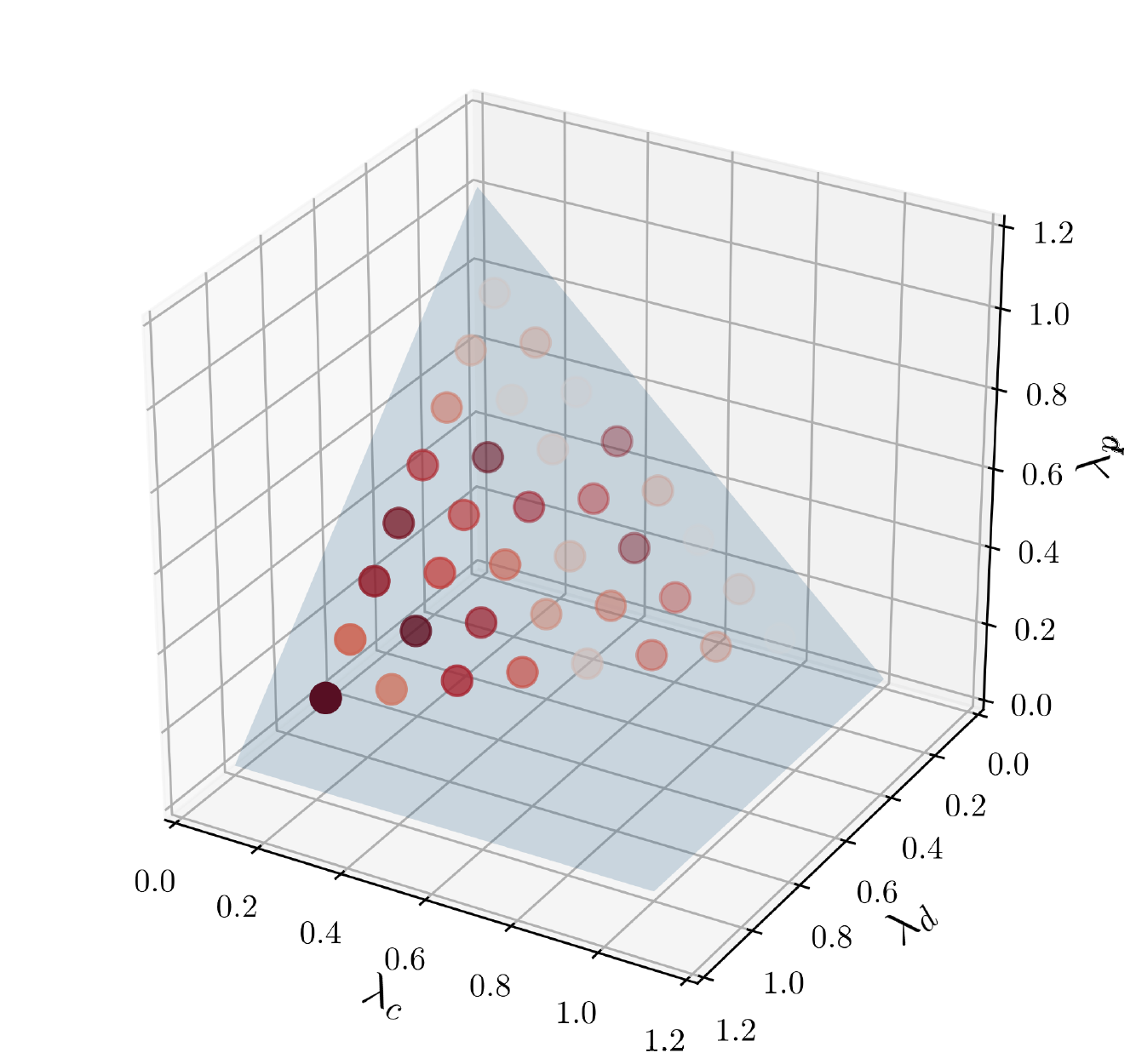}}
		\vspace{-0.3cm}
	\captionsetup{justification=centering}
	\caption{Reward Analysis {\it w.r.t.} Tokyo dataset.}
		\vspace{-0.3cm}
	\label{fig:tky_reward}
\end{figure*}

\subsection{Robustness Check (\textbf{Q2})}
We randomly divided the dataset into 5 subsets, and evaluated DRPR over these subsets to examine its robustness (low variance). 
As illustrated in Figure \ref{fig:nyc_robust} and Figure \ref{fig:tky_robust}, we can find that compared with our preliminary framework RIRL, the enhanced framework DRPR is more stable in terms of Prec\_Cat, Rec\_Cat, Avg\_Sim, and Avg\_Dist.
Such an observation indicates that the model performance is more robust as capturing more user mobility patterns.
In addition, another interesting observation is that DRPR outperforms RIRI under the same sampling strategy.
Such an observation reflects that capturing connected, topological, and semantic interactions make DRPR have a more stable performance compared with RIRL.
Moreover, after a careful inspection of Figure \ref{fig:nyc_robust} and Figure \ref{fig:tky_robust} , we can find that no matter in DRPR or RIRL, the TD-based sampling strategy is more superior to the reward-based sampling strategy.
A possible interpretation for the observation is that the TD-based sampling strategy enhances the comprehension of the reinforced agent for the data samples that are difficult to learn.

\vspace{-0.2cm}

\subsection{Ablation Studies for DRPR (Q3)}
We conducted ablation studies for DRPR to validate the necessity of each technical component.
The model variants are: 1)
 DRPR$^*$, which replaces the dynamic KG of DRPR with the static KG.
 2) DRPR$^\prime$, which removes the exit mechanism of DRPR.
 3) DRPR$^-$, which removes the POI candidate generation part.
 Table \ref{tab:ablation_study_nyc} and Table \ref{tab:ablation_study_tky} show the comparison results in terms of all evaluation metrics.
 We can find that DRPR outperforms DRPR$^*$ on both New York and Tokyo datasets.
This observation indicates that
the dynamic updating process in the dynamic KG can not only capture the temporal patterns of the environment, but also grasp the change in user mobility preference accurately.
Additionally, the performance of DRPR$^\prime$ drops significantly compared with DRPR. 
A possible reason is that the exit mechanism discards the outdated and trivial user mobility interests, avoiding the learning  of DRPR from being disturbed by them.
Moreover, DRPR exhibits a great improvement on both recommendation performance and convergence speed on both New York and Tokyo Datasets. 
A potential reason is that the personalized POI candidate set  significantly reduces the action space for the agent, making the exploration on POI more efficient.
Meanwhile, as the personalized POI candidate is more accurate and targeted for the user, the recommendation effectiveness can also be improved.
In summary, this experiment validates that each technical component is indispensable and significant.

\subsection{The Study of Reward Function (\textbf{Q4})}
Our reward function consists of three components, i.e., distance $r_d$, category similarity $r_c$, and prediction accuracy $r_p$ whose value is 1 if the predicted POI is identical to the real user visit event, and 0 otherwise. 
We combined the three components into  with three  weights, i.e., $\lambda_d$, $\lambda_c$ and $\lambda_p$, {\it w.r.t} $\lambda_d + \lambda_c + \lambda_p = 1$.
We set the learning rate as 1$e$-5, and project evaluation metrics into a 3D space, with ($r_d$, $r_c$,$r_p$) as axes. The shade of color denotes the value of the metric.
The darker the color is, the higher the performance will be.

 Figure \ref{fig:nyc_reward} and Figure \ref{fig:tky_reward} show an interesting observation: the contribution of $r_d$ is higher than $r_c$ and $r_p$ in terms of all evaluation metrics.
The reason is that we have employed the meta-path-based POI candidate generation in DRPR to discover the POI and POI category subset that a given user is possibly interested in.
This operation improves the importance of $r_d$: as long as DRPR predicts the POIs that are close to real visit POIs, the model performance can be improved from all sides. 
A careful inspection of Figure \ref{fig:nyc_reward} and Figure \ref{fig:tky_reward} find  that although $r_d$ is more important than the other two factors for all metrics, the best model performance is achieved based on the balanced combination of $r_d$, $r_c$, and $r_p$.
This observation reveals that balanced considering different factors can improve the comprehension of our model for user preferences.

\section{Related Works}
\noindent{\bf POI Recommendation.}
POI recommendation plays an important role in location-based social networks (LBSNs).
Accurate POI recommendations help users explore interesting places, which makes their lives more convenient.
Thus, many researchers are attracted to the POI recommendation domain
~\cite{wang2018you,liu2018incorporating,wang2019adversarial,liu2018modeling,wang2019spatiotemporal}.
For example, Lian {\it et al.} ~\cite{lian2014geomf} explored user mobility preferences by a modified weighted matrix factorization method.
Liu {\it et al.} proposed a novel geographical probabilistic factor analysis framework to study the user geographical interest \cite{liu2013learning}.
Compared with these works,  DRPR incorporates multiple factors that affect POI recommendation into a dynamic knowledge graph, and the exploration and exploitation modules of the reinforced agent in DRPR capture user visit preferences sufficiently.

\noindent{\bf Knowledge Graph-based Recommendation.}
Knowledge graph (KG) demonstrates the semantic relations and reasoning structure among different entities.
Owing to the rich semantic information in KG,
many researchers adopt KG in the recommendation domain and achieve  good performance ~\cite{zhang2016collaborative,sun2018recurrent,wang2020reinforced,chen2020jointly}. 
For instance, Sun {\it et al.} utilized a recurrent neural network to learn the semantic rich embedding of entities and relations for capturing the user preference ~\cite{sun2018recurrent}.
Xian {\it et al.} proposed the PGPR framework that employs the meta-path in KG for explainable recommendations ~\cite{xian2019reinforcement}.
Wang {\it et al.} generated the path embedding by incorporating the semantics of both entities and relations and inferred reasonable recommendation based on meta-paths ~\cite{wang2019explainable}.
Compared with the previous works, we implement a dynamic KG to simulate the environment where users visit.
Furthermore, we unify the dynamic KG's embedding and meta-paths to sufficiently explore user mobility preferences.

\noindent{\bf Reinforcement Learning for Online Recommendation.}
Reinforcement learning-based recommendation systems formulate the user-item interaction into a sequential decision-making process for recommending items ~\cite{wang2018reinforcement,zheng2018drn,he2020learning,liu2020ambulance}. 
For instance, Zheng {\it et al.}  utilized reinforcement learning to capture implicit user feedback characteristics ~\cite{zheng2018drn} for news recommendation.
POI recommendation is different from traditional recommendation system, because its strong geographical constrains.
Recently, in POI recommendation domain, Wang {\it et al.} incorporated reinforcement learning and a spatial knowledge graph to grab user mobility pattern ~\cite{wang2020incremental}.
Compared with ~\cite{wang2020incremental}, we leverage a dynamic KG to incorporate the interaction and information of users and a geographical environment.
Meanwhile, we recommend attractive POIs for users by exploring the substructure information in the dynamic KG.

\section{Conclusion Remarks}
In this paper, we propose a novel POI recommendation framework by integrating dynamic KG into a reinforcement learning setting.
Since the common practice of KG-based recommendation cannot capture the multi-level dynamics of human mobility, we introduce dynamic KG as the environment.
Specifically, we regard the representation of the dynamic KG as the state in reinforcement learning.
To address the uncontrollable graph-scale issue, we develop an exit mechanism to remove outdated information.
To reduce the action space, we devise a meta-path-based POI candidate generation method to select the most possible POIs to recommend.
To solve the problem of the dynamic action space, we propose a new policy network that takes state-action pairs as input and outputs a Q value for each pair.
We select the pair with the highest Q value as the recommendation.
From extensive experiments, we can find that our proposed method can better capture the dynamic characteristics of user mobility patterns and preferences compared with other baselines.
And the personalized POI candidates make the recommendation performance more accurate and efficient.


\section*{Acknowledgments}
This research was supported by the National Science Foundation(NSF) via the grant number: 2040950, 2006889, 2045567.

%
\bibliographystyle{plain}
\bibliography{tbd}

\vspace{-30pt}

\begin{IEEEbiography}
[{\includegraphics[width=1in,height=1.25in,clip,keepaspectratio]{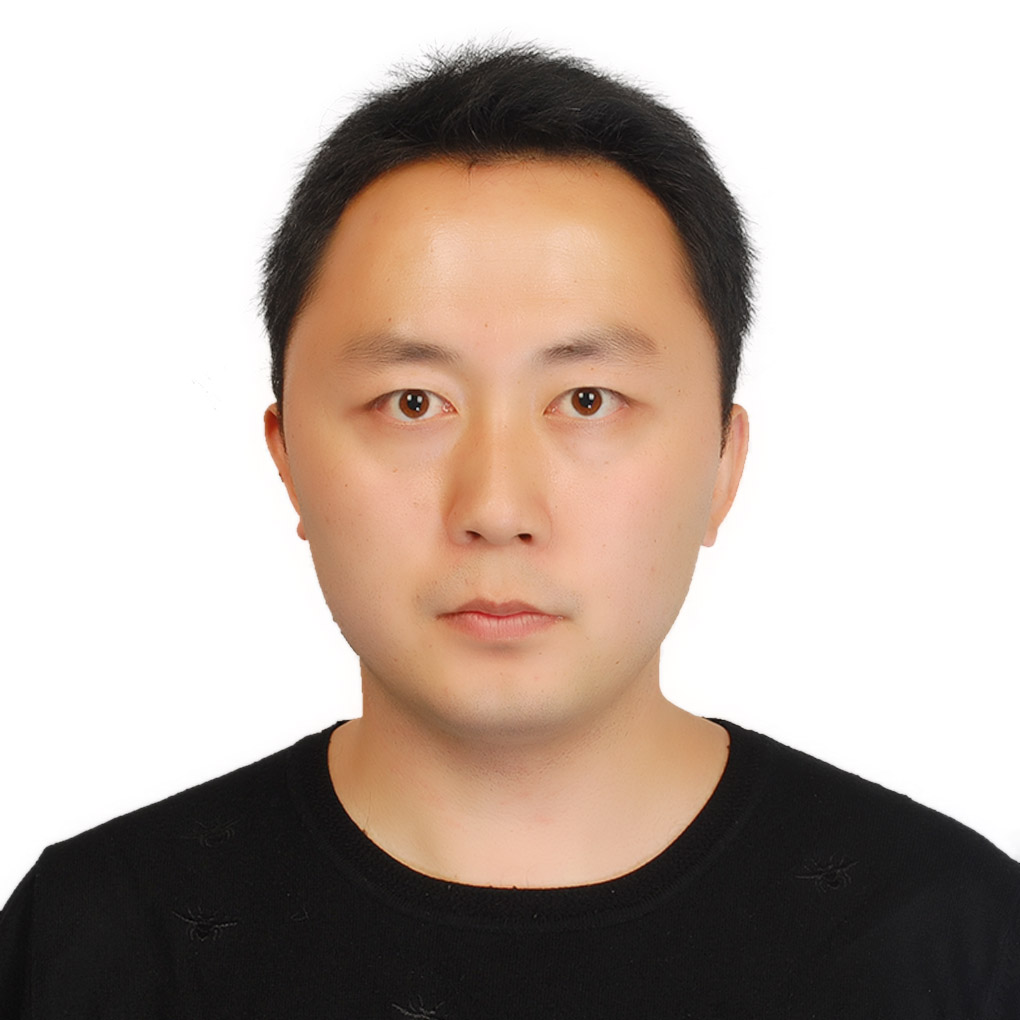}}]
{Dongjie Wang}
 received the BE degree from
the Sichuan University, China, 2016, the MS degree from the Southwest Jiaotong University, China, 2017. He is currently working toward the PhD degree at the University of Central Florida (UCF). His research
interests include Anomaly Detection, Generative Adversarial Network, Reinforcement Learning and Spatio-temporal Data Mining.
\end{IEEEbiography}

\vspace{-20pt}
\begin{IEEEbiography}[{\includegraphics[width=1in,height=1.25in,clip,keepaspectratio]{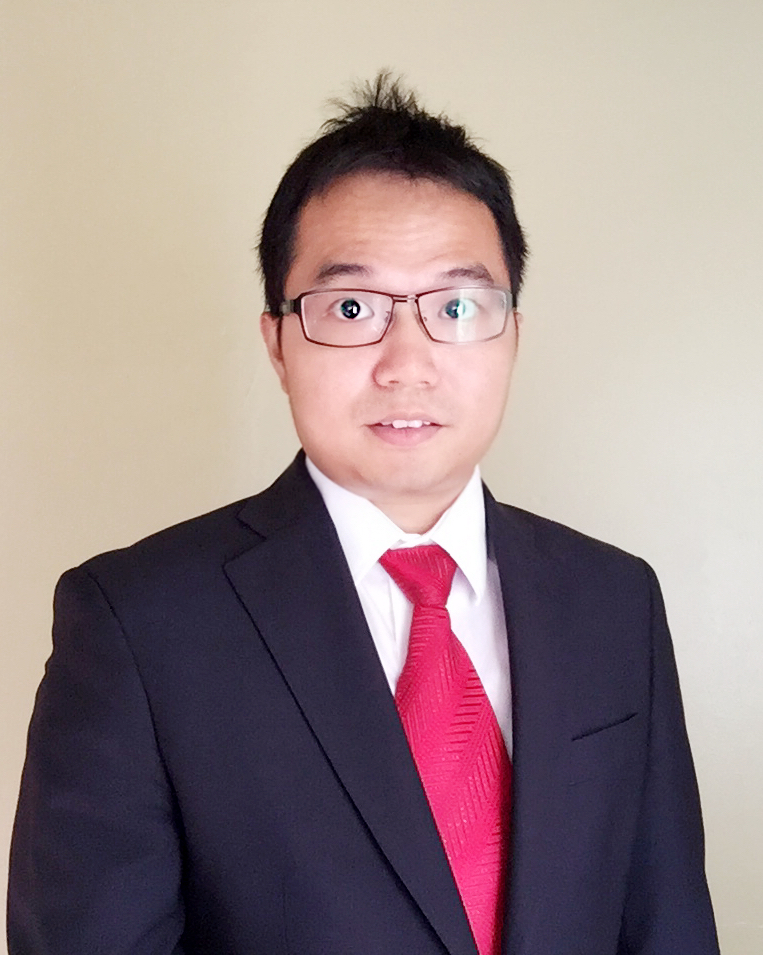}}]{Yanjie Fu}
is an Assistant Professor in the Department of Computer Science at the University of Central Florida. He received his Ph.D. degree from Rutgers, the State University of New Jersey in 2016, the B.E. degree from University of Science and Technology of China in 2008, and the M.E. degree from Chinese Academy of Sciences in 2011. His research interests include Data Mining and Big Data Analytics.
\end{IEEEbiography}

\vspace{-20pt}
\begin{IEEEbiography}
[{\includegraphics[width=1in,height=1.25in,clip,keepaspectratio]{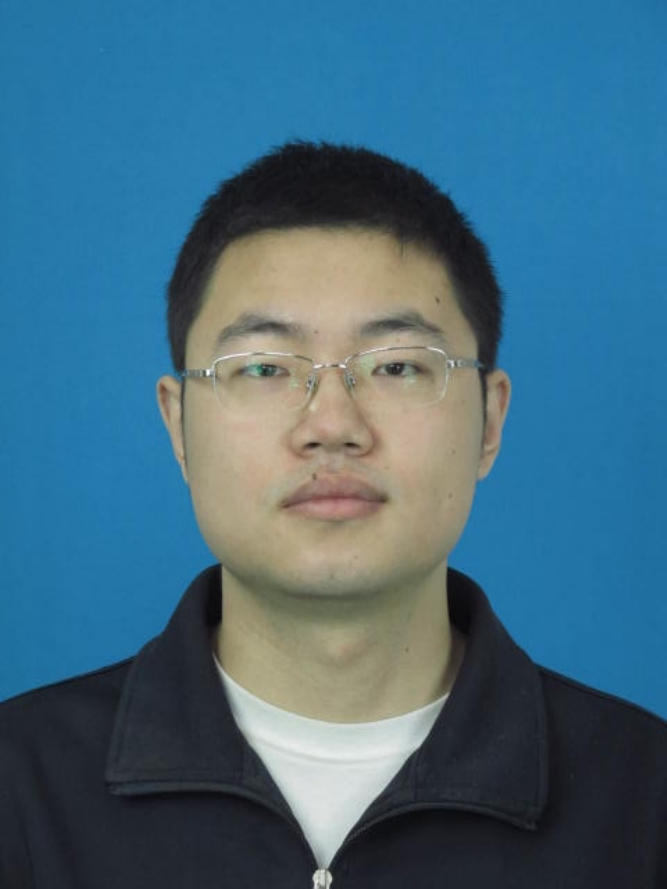}}]
{Kunpeng Liu} is an Assistant Professor in the Department of Computer Science, Portland State University. Prior to that, he received the Ph.D. degree from the CS Department of the University of Central Florida. He received both his M.E. degree and B.E. degree from the Department of Automation, University of Science and Technology of China (USTC). He has rich research experience in industry research labs, such as Geisinger Medical Research, Microsoft Research Asia, and Nokia Bell Labs. He has published in refereed journals and conference proceedings, such as IEEE TKDE, ACM SIGKDD, IEEE ICDM, AAAI, and IJCAI. He has received a Best Paper Runner-up Award from IEEE ICDM 2021.
\end{IEEEbiography}

\vspace{-20pt}

\begin{IEEEbiography}[{\includegraphics[width=1in,height=1.25in,clip,keepaspectratio]{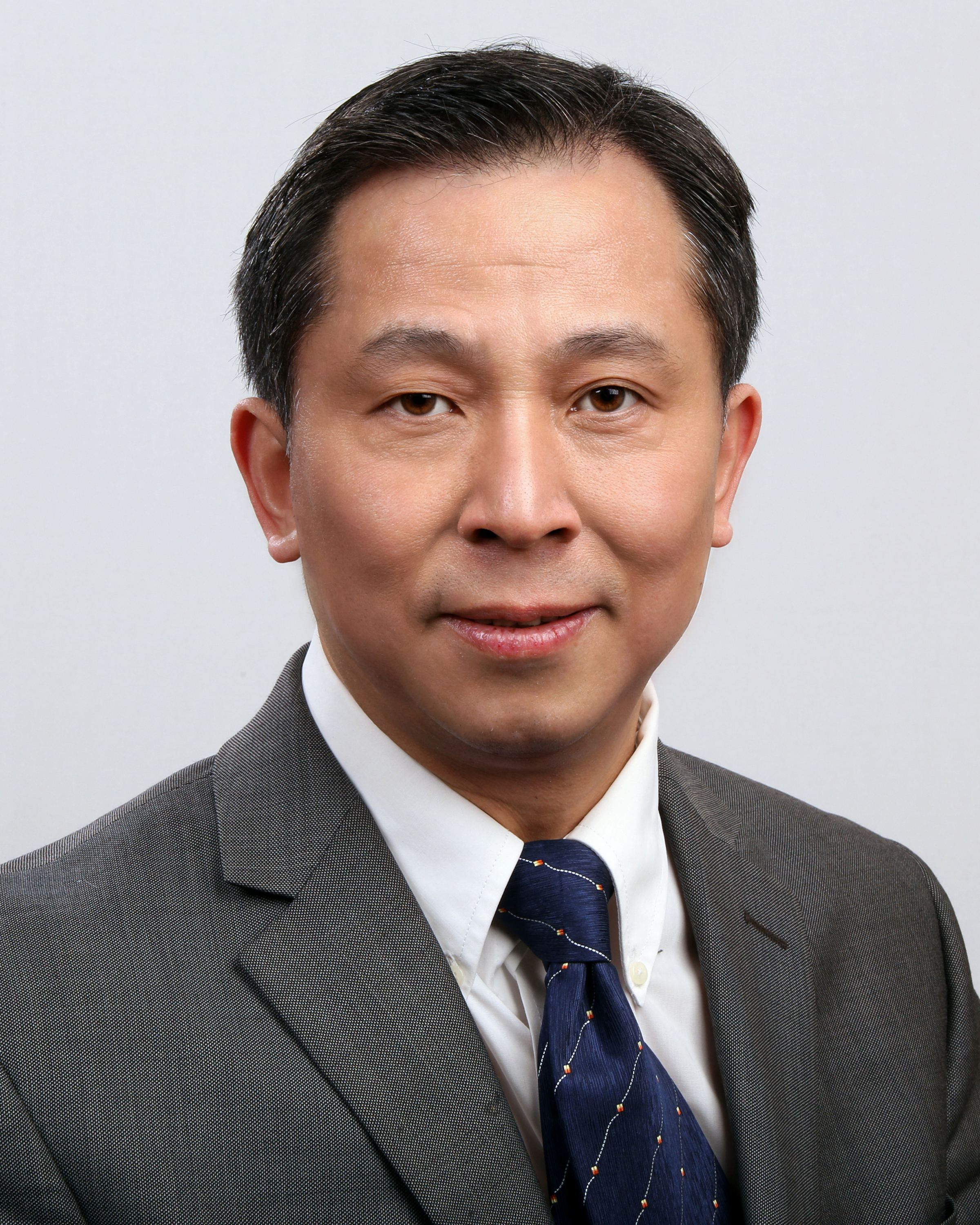}}]{Hui Xiong}
 is currently a Chair Professor at the
Hong Kong University of Science and Technol-
ogy (Guangzhou). Dr. Xiong’s research interests
include data mining, mobile computing, and their
applications in business. Dr. Xiong received his
PhD in Computer Science from University of
Minnesota, USA. He has served regularly on the
organization and program committees of numer-
ous conferences, including as a Program Co-
Chair of the Industrial and Government Track for
the 18th ACM SIGKDD International Conference
on Knowledge Discovery and Data Mining (KDD), a Program Co-Chair
for the IEEE 2013 International Conference on Data Mining (ICDM), a
General Co-Chair for the 2015 IEEE International Conference on Data
Mining (ICDM), and a Program Co-Chair of the Research Track for the
2018 ACM SIGKDD International Conference on Knowledge Discovery
and Data Mining. He received the 2021 AAAI Best Paper Award and
the 2011 IEEE ICDM Best Research Paper award. For his outstanding
contributions to data mining and mobile computing, he was elected an
AAAS Fellow and an IEEE Fellow in 2020.
\end{IEEEbiography}





\end{document}